\documentclass[twocolumn,showpacs,floatfix,amsmath,amssymb,prb,superscriptaddress]{revtex4}
\date{\today}

\usepackage{color} 

\usepackage{graphicx}% Include figure files
\usepackage{amsmath}
\usepackage{amssymb}

\newcommand{\mrm}[1]{\mathrm{#1}}

\def\e{\mathrm{e}}
\def\I{\mathrm{i}}

\newcommand{\be}{\begin{equation}}
\newcommand{\ee}{\end{equation}}
\newcommand{\ba}{\begin{eqnarray}}
\newcommand{\ea}{ \end{eqnarray}}

\begin{document}

%\def\openone{\leavevmode\hbox{\small1\kern-4.2pt\normalsize1}}
%%%%%%%%%%%%%%%%%%%%%%%%%%%%%%%%%%%%%%%%%%%%%%%%%%%%%%%%%%%%%%%%%%%%%%%%%%%%%%%
\title{Vibrational cooling and thermoelectric response of nanoelectromechanical systems}

\author{Liliana Arrachea} 
\affiliation{Departamento de F\'{\i}sica, Facultad de Ciencias Exactas y Naturales and IFIBA, Universidad de Buenos Aires, Pab.\ I, Ciudad Universitaria, 1428 Buenos Aires, Argentina }
 
\author{Niels Bode} 
\affiliation{\mbox{Dahlem Center for Complex Quantum Systems and Fachbereich Physik, Freie Universit\"at Berlin, 14195 Berlin, Germany}}

 \author{Felix von Oppen}
\affiliation{\mbox{Dahlem Center for Complex Quantum Systems and Fachbereich Physik, Freie Universit\"at Berlin, 14195 Berlin, Germany}}

\date{\today}
\begin{abstract}
An important goal in nanoelectromechanics is to cool the vibrational
motion, ideally to its quantum ground state. Cooling by an applied charge current
is a particularly simple and hence attractive strategy to this effect.
Here, we explore this phenomenon in the context of the general theory of
thermoelectrics. In linear response, this theory describes thermoelectric
refrigerators in terms of their cooling efficiency $\eta$ and figure of
merit $ZT$. We show that both concepts carry over to phonon cooling in
nanoelectromechanical systems. As an important consequence, this allows us
to discuss the efficiency of phonon refrigerators in relation to the
fundamental Carnot efficiency. We illustrate these general concepts by
thoroughly investigating a simple double-quantum-dot model with the dual
advantage of being quite realistic experimentally and amenable to a
largely analytical analysis theoretically. Specifically, we obtain results
for the efficiency, the figure of merit, and the effective temperature of
the vibrational motion in two regimes. In the quantum regime in which the
vibrational motion is fast compared to the electronic degrees of freedom,
we can describe the electronic and phononic dynamics of the model in terms
of master equations. In the complementary classical regime of slow
vibrational motion, the dynamics is described in terms of an appropriate
Langevin equation. Remarkably, we find that the efficiency can approach
the maximal Carnot value in the quantum regime, with large associated
figures of merit. In contrast, the efficiencies are typically far from the
Carnot limit in the classical regime. Our theoretical results should provide guidance
to implementing efficient vibrational cooling of nanoelectromechanical
systems in the laboratory.
\end{abstract}
\maketitle

\section{Introduction}
Cooling nanomechanical systems into the quantum ground state has attracted much attention for some years now. One of the most explored mechanisms is the
coupling of the nanomechanical motion to photons in optical cavities. Inspired by the laser cooling of Bose-Einstein condensates and cold atoms, this  technique enabled the observation of features related to the quantum zero-point fluctuations of a mechanical device.\cite{Safavi2012,mar-gir} 

The electron-phonon coupling offers an alternative route towards cooling a nanomechanical system. This mechanism has the appealing property that it is operated simply by a bias voltage driving an electronic current through a suitably engineered nanoelectromechanical structure (NEMS). Exploiting the electron-phonon interaction to refrigerate a NEMS with a {\em dc} current has been studied in several theoretical works,
\cite{zippilli-2009,zippilli-2010} focusing on carbon nanotubes \cite{gorelik-2011} as well as molecular setups.\cite{pistolesi-2009,galperin-2009,mceniry-2009,galptef} The principal ingredient is an asymmetry in the operation of the device that favors the absorption over the emission of phonons. The underlying processes exhibit some similarity to electron cooling as implemented experimentally in quantum dots and micrometer-scale electronic systems containing normal and superconducting pieces.\cite{ naik-2006,prance-2009, muhonen-2012} Yet another approach to refrigerating a nanomechanical system which we will not consider here uses pumping of phonons.\cite{phon} 

\begin{figure}[tb]
  \begin{center}
      \includegraphics[width=5cm]{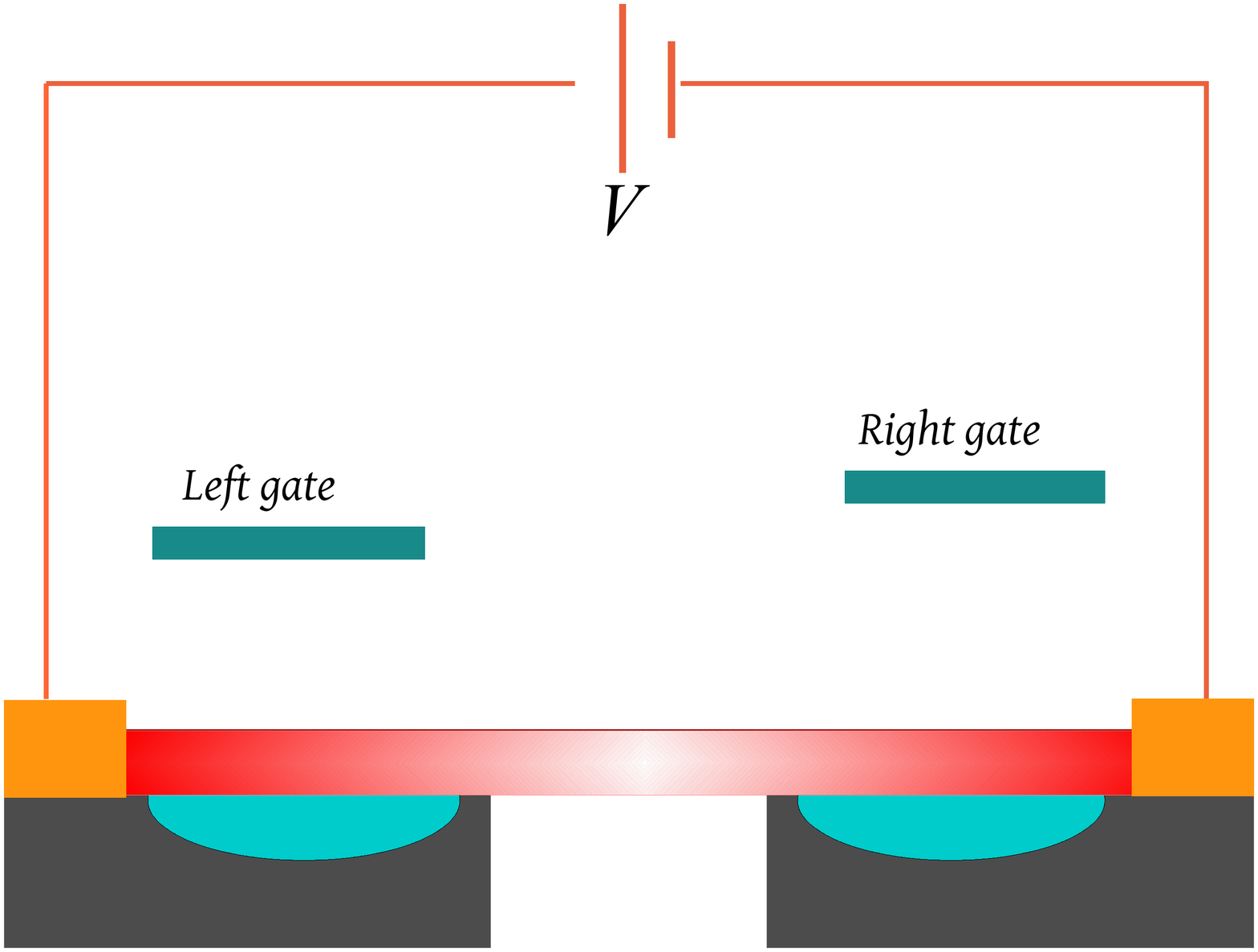}
        \includegraphics[width=4.7cm]{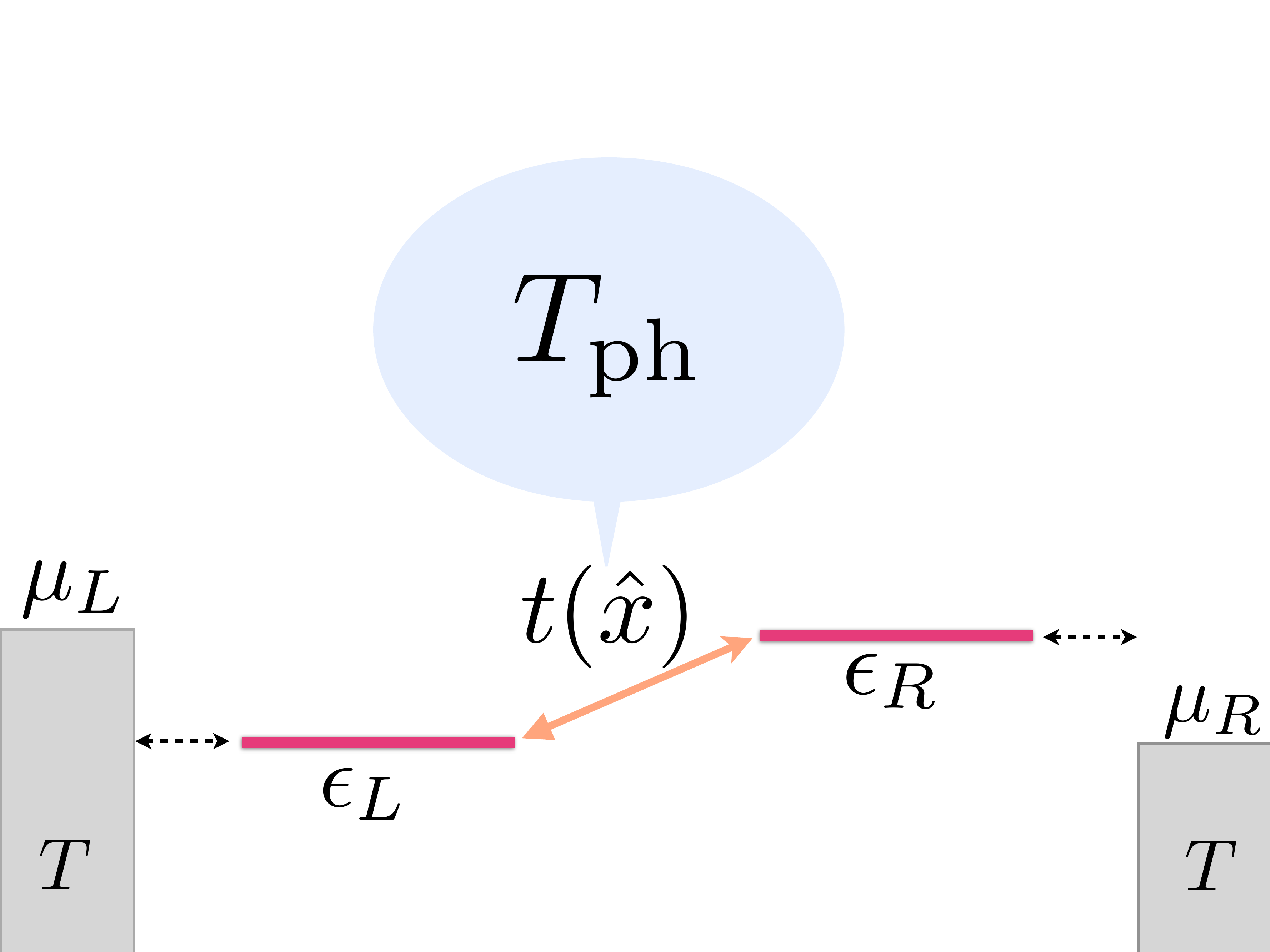}
  \end{center}
\caption{(Color online) Top: Sketch of the setup under consideration. A suspended nanostructure (e.g., a carbon nanotube) is contacted by left and right electrodes. The setup consists of two gate-tunable quantum dots. Tunneling between the quantum dots couples to vibrational motion of the 
suspended part of the structure. Bottom: Configuration for vibrational cooling. The electronic levels $\epsilon_L$ and $\epsilon_R$ of the quantum dots 
are arranged such that electron tunneling between the dots is preferentially accompanied by phonon absorption, enabled by the dependence of the tunneling amplitude $t(\hat x)$ on the vibrational mode coordinate $\hat x$. We also consider the situation when the system is in contact with an additional phonon bath of temperature $T_{\rm ph}$.}
\label{fig:setup}
\end{figure}

Our present study of electronic cooling of nanomechanical motion is motivated by two recent developments. Theoretically, there has been increasing interest in thermoelectrics for quantum nanostructures.\cite{casati1,casati2,seifert,janine,ora1,ora2} The basic approach typically relies on linear-response relations for both charge and heat currents and uses general principles such as the second law of thermodynamics to derive relations between these coefficients as well as bounds on the efficiency of thermoelectric machines. Previous works on the electronic cooling of nanomechanical motion have not made connection with this general theory. As we will show, establishing this connection explicitly allows one to describe phonon refrigerators in terms of the efficiencies and figures of merit which are customary in thermoelectrics. Experimentally, there have been significant advances in controlling the electronic and phononic structure as well as their interaction in suspended carbon nanotubes.\cite{chaste,fel-dot} These advances should have significantly reduced the engineering challenges in realizing some of the cooling devices discussed in the literature. In this paper, we discuss the refrigeration device sketched in Fig.\ \ref{fig:setup}. It consists of two gate-tunable quantum dots coupled by a suspended tunnel junction. Electrons tunneling between the quantum dots can excite the (for simplicity: single) vibrational mode of the suspended section. Besides its realism, this model has the added benefit that it is amenable to an essentially analytical description if we assume that the tunneling between the dots is weak compared to their couplings to the leads. A similar model was considered in Ref.\ \onlinecite{lu-2011} to describe rectification in molecular junctions. 

This device operates as a refrigerator for the mechanical motion when the dot levels increase in energy in the direction of current flow.\cite{zippilli-2009,zippilli-2010}  In this setting, electron transport is preferentially accompanied by the absorption of phonons and thus causes phonon refrigeration. We explore the efficiency of this cooling mechanism in two regimes which we refer to as quantum and classical. In the quantum regime, the tunneling amplitude between the two dots is weak so that the mechanical frequency is large compared to the rate at which electrons are passing through the structure. In this regime, the cooling can be thought of in terms of phonon-assisted tunneling of the electrons and adequately described in terms of a rate equation. \cite{koch-2006,pistolesi-2009} This rate equation allows one to calculate the non-equilibrium phonon distribution as well as the general linear response coefficients entering the general theory of thermoelectric response. In the complementary classical regime, the mechanical oscillations are slow compared to the rate at which electrons are passing through the structure. In this limit, the mechanical motion can be treated in terms of a classical Langevin equation, with the electron-phonon interaction accounted for in terms of effective forces including a fluctuating force.\cite{pisto-class,bode-2011,bode-2012} Cooling has not yet been investigated theoretically within this classical regime, although this regime is actually important in several recent experiments.\cite{chaste,fel-dot}  Specifically, experiments on suspended carbon nanotubes can be performed in both the quantum and the classical regime. However, flexural modes of suspended carbon nanotubes typically have low frequencies, often requiring a classical description.  

To characterize the refrigeration device, we consider two setups. In one setup, we assume that the mechanical motion is strongly coupled to a phonon reservoir with fixed temperature $T_{\rm ph}$ (see also Refs.\ \onlinecite{ora1} and \onlinecite{ora2}). This is appropriate when the coupling of the mechanical motion to the phonon reservoir (i.e.,  to non-electronic degrees of freedom) causes faster relaxation processes than the coupling to the electrons. The cooling strength of the device can then be characterized in terms of the heat current that is extracted from the phonon reservoir. The associated efficiency is defined as the ratio of the extracted heat current and the power invested in the electron system. We evaluate this efficiency within linear response, valid when the electron temperature $T$ is not too different from the phonon temperature $T_{\rm ph}$ and compare it to the maximal Carnot efficiency. In another setup, we assume that the phonon motion is entirely controlled by the coupling to the electrons. This is appropriate when the coupling of the mechanical motion to a reservoir is sufficiently weak or entirely absent. The cooling power is now characterized by the effective temperature of the mechanical motion, defined through the condition that no heat current flows between the mechanical mode and a (fictitious) weakly-coupled phonon reservoir.\cite{en-an,leto1,leto2,lili,hugo,eftem} Within linear response, this effective temperature can also be obtained directly from the general linear-response coefficients. 

This article is organized as follows. Sec.\ \ref{maqt} introduces the model by which we describe the device in Fig.\ \ref{fig:setup} (Sec.\ \ref{sec:cooling_model}) and briefly summarizes essential results of the theory of the thermoelectric response (Sec.\ \ref{thermoelectric}). In Sec.\ \ref{sec:quantum_regime}, we consider the quantum regime of fast mechanical motion. Sec.\ \ref{sec:classical_regime}  discusses the complementary classical regime of fast electronic dynamics. We summarize and conclude in Sec.\ \ref{sec:conclusions}. Some details of the calculations are relegated to appendices.

\section{Model and thermoelectric response}
\label{maqt}

\subsection{Model} 
\label{sec:cooling_model}

The two-quantum-dot setup described above and depicted in Fig.\ \ref{fig:setup} can be modeled by the Hamiltonian
\be \label{ham}
H= H_{\rm el}+ H_{\mathrm{T}}+ H_{\mathrm{v}} + H_{{\rm ph}}+ 
H_{c, {\rm ph}}.
\ee
Here, the first term accounts for the two quantum dots and their couplings to the two electrodes, 
\be \label{hamel}
H_{\rm el}=
\sum_{\alpha=L,R} (H_{\alpha} + H_{c,\alpha}  + \epsilon_{\alpha} d^{\dagger}_{\alpha} d_{\alpha} ).
\ee
Both quantum dots $\alpha=L,R$ host one electronic state of energy $\epsilon_{\alpha}$, with corresponding creation (annihilation) operators $d^{\dagger}_{\alpha}$ ($d_{\alpha}$). The dots are assumed noninteracting and in contact with one electron reservoir each. The reservoirs are modeled by the free-electron Hamiltonians
\be
H_{\alpha}=\sum_{k_{\alpha} } \epsilon_{k_{\alpha}} c^{\dagger}_{k_{\alpha} }c_{k_{\alpha} },
\ee
where $c^{\dagger}_{k_{\alpha} }$ ($c_{k_{\alpha} }$) creates (annihilates) an electron in state $k_{\alpha}$ of electrode $\alpha$ and the 
hybridization between quantum dots and electrodes is described by
\be
H_{c,\alpha}= \sum_{k_{\alpha}} w_{k_{\alpha}} c^{\dagger}_{k_{\alpha}} d_{\alpha} + \mrm{h.c.}
\ee
with the amplitude $w_{k_{\alpha}}$. 

The vibrational mode couples to the electronic degrees of freedom through the tunnel coupling between the two quantum dots, 
\begin{align} \label{interdot}
  H_{\mathrm{T}}=t(\hat{x})  d^{\dagger}_{L} d_R + \mrm{h.c.}\, .
\end{align}
Specifically, the tunneling amplitude $t(\hat{x}) = t_0 \e^{- \lambda \hat{x}}$ depends on the vibrational coordinate $x$, which provides the electron-phonon coupling of strength $\lambda$.\cite{koch-2004} 
For simplicity, we assume that the mechanical motion is characterized by a single normal-mode coordinate. Expressing this coordinate in terms of phononic creation and annihilation operators, $\hat{x} = \hat{a} + \hat{a}^\dagger$, the free motion of the vibrational mode is governed by the Hamiltonian
\be
H_{\mathrm{v}}= \omega (a^{\dagger} a+ \frac{1}{2}) ,
\ee
where $\omega$ is the frequency. 

Finally, the last two terms of the Hamiltonian (\ref{ham})
represent a phonon bath and its coupling to the vibrational mode. We will provide some further details for these contributions in Sections III and IV.

For the most part of the manuscript, we will set $\hbar=k_B=1$ unless a restoration of conventional units facilitates the discussion. 

\subsection{Thermoelectric description}
\label{thermoelectric}

We briefly review some aspects of the general theory of thermoelectric response\cite{casati2} in a form adapted to the refrigeration of a vibrational mode. In many ways, our discussion here follows Refs.\ \onlinecite{ora1} and \onlinecite{ora2} which consider a three-terminal setup including two electron reservoirs and one phonon reservoir. 

We focus attention on thermoelectric cooling of the vibrational mode coupled to a phonon reservoir at temperature $T_{\rm ph}$ and assume that the two electron reservoirs are at the same temperature $T$. Charge currents $J^C$ between the two electron reservoirs and heat currents $J^Q$ from the phonon to the  electron reservoirs can be induced by applying a chemical potential difference $\Delta \mu = \mu_L - \mu_R$ between the electron reservoirs or a temperature difference $\Delta T = T_{ph} - T$. Within linear response, the thermoelectric effects are then described in terms of a $2 \times 2$ matrix
 ${\bf L}$, 
\[ \left( \begin{array}{c}
J^C/e \\
J^Q \end{array} \right)= \left(\begin{array}{cc}
L_{11} & L_{12} \\
L_{21} & L_{22}  \end{array} \right) 
 \left( \begin{array}{c}
\Delta\mu/T \\
\Delta T/ T^2 \end{array}
\right), \]
or in short ${\bf J}= {\bf L} \cdot {\bf X}$. Here, the quantities $X_1 = \Delta\mu/T$ and $X_2 = \Delta T/T^2$ are known as affinities. The Onsager reciprocity relations yield $L_{12}(B)=L_{21}(-B)$ in the presence of a magnetic field $B$. From now on, we will assume that the system is time-reversal symmetric so that $L_{12}=L_{21}$.

Our device operates as a refrigerator as long as $J^Q>0$ for $T_{ph}<T$. Given a certain bias voltage $V = \Delta\mu/e$, this is the case for phonon temperatures in the interval $ T(1-L_{21}\Delta\mu/L_{22}) < T_{\rm ph} < T$.

We can characterize the operation of the device in Fig.\ \ref{fig:setup} as a refrigerator through the coefficient of performance $\eta$, which is defined as the ratio of the rate at which heat is extracted from the cold reservoir (i.e.\ the phonon reservoir) and the invested electric power,
\be \label{eta}
\eta= \frac{\dot{Q}}{\dot{W}}= \frac{J^Q}{(J^C/e)\Delta \mu}= \frac{L_{21}X_1+L_{22}X_2}{T X_1 (L_{11} X_1 + L_{12}X_2)}.
\ee
This efficiency can be related to the rate of entropy production, $\dot{\cal S} = (J^C/e) X_1 + J^Q X_2$ which yields
\begin{equation}
   \eta = \eta_C\left( 1 - \frac{T\dot{\cal S}}{(J^C/e)\Delta\mu}\right).
\end{equation}
Thus, as a consequence of the second law of thermodynamics, the efficiency $\eta$ is always smaller than the Carnot efficiency for refrigeration (given here to linear-response accuracy),
\begin{equation}
  \eta_C = \frac{T}{|\Delta T|}.
\end{equation}
We also note another consequence of the second law. Writing the rate of entropy production in linear response,
\be
\dot{\cal S}= {\bf X}^{\rm t} \cdot {\bf L} \cdot {\bf X},
\ee
we conclude that ${\bf L}$ is positive semidefinite, i.e.\
\ba
& & L_{11},\;L_{22} >0, \nonumber \\
& & L_{11} L_{22} - L_{12}^2 \geq 0.
\ea
In addition to the currents $J^C$ and $J^Q$, there will also be a heat current flowing between the two electron reservoirs in our device. However, this current does not contribute to entropy production as it flows between two reservoirs of equal temperature. More generally, it does not play an essential role in the following.

We can also define a figure of merit $ZT$ for our three-terminal setup in the usual manner. Indeed, for a given temperature difference $\Delta T$, the efficiency can be maximized as function of voltage. This yields the maximal efficiency 
\be
  \eta=\eta_C \; \frac{\sqrt{1+Z T }-1}{\sqrt{1+Z T}+1},
\ee
where 
\be \label{z}
Z T = \frac{L_{12}^2}{  \mbox{det}\left({\bf L}\right)},
\ee
is the figure of merit. Thus the Carnot efficiency would be attained as $ZT \rightarrow \infty$. 

So far, we assumed that the vibrational mode is coupled to a phonon reservoir which fixes its temperature to $T_{\rm ph}$. Alternatively, we could consider the vibrational mode decoupled from the phonon reservoir. In this case, cooling can be characterized by an effective temperature of the vibrational mode which is smaller than the electron temperature, as done in previous works.

For a general nonequilibrium situation, the distribution function of the vibrational mode will not be thermal so that we need to specify what we mean by effective temperature. A possible definition in a non-equilibrium transport setup relies on coupling the vibrational mode to a thermometer, a reservoir with infinitesimal coupling to the vibrational mode.\cite{en-an,hugo,lili,eftem} The effective temperature is then defined as the temperature at which there is vanishing heat flow between thermometer and vibrational mode. This definition was originally introduced by Engquist and Anderson \cite{en-an} and has been widely adopted in many transport setups. This definition allows us to obtain the effective temperature of the vibrational mode within the above formalism by requiring that $J^Q=0$ which yields
\be \label{tloc}
T^{\rm eff}= T\left( 1-eV \frac{L_{12}}{L_{22}}\right).
\ee
Note that this is just the minimal phonon temperature at which the device with a phonon thermostat operates as a phonon refrigerator. 

\section{Quantum regime} 
\label{sec:quantum_regime}

We first consider the quantum regime in which the tunneling rate between the quantum dots is small compared to the vibrational frequency. Moreover, we assume that the coupling between quantum dots and leads is strong compared to the coupling between the quantum dots. In this limit, we can describe the system in terms of a master (or rate) equation for the occupation probability $P_{n}$ of the phonon mode. Here, $P_n$ denotes the probability that the phonon state of energy $n\omega$ is occupied. The state of the phonon mode can change whenever an electron tunnels between the two quantum dots and the corresponding rates can be readily derived from Fermi's Golden Rule.

\subsection{Rate equation}

We first set up the master equation for the dynamics of the phonon population. Following Ref.\ \onlinecite{koch-2006}, the master equation for  $P_n$  takes the form 
 \begin{align}
\dot{P_n}=&-P_n \sum_{n^{\prime} } W^{n \rightarrow n^{\prime}} + \sum_{n^{\prime}} P_{n^{\prime}} W^{n^{\prime} \rightarrow n } - \frac{1}{\tau} [P_n -P_n^{\mrm{eq}}], 
\label{eq:rateequation}
\end{align}
where $W^{n \rightarrow n^{\prime}}$ denotes the rate of transitions from phonon state $n$ to $n^{\prime}$. The last term in Eq.\ (\ref{eq:rateequation}) accounts for the coupling of the oscillator to the phononic environment in a phenomenological manner.\cite{koch-2006} 
We assume that this phonon heat bath is at a temperature $T_{\rm ph}$, so that the phonon distribution $P_n$ will relax to the equilibrium distribution 
\be \label{peq}
P_n^{\mrm{eq}}=\mrm{e}^{-n  \omega/T_{\rm ph}}(1-\mrm{e}^{-\omega/T_{\rm ph}}),
\ee
within the relaxation time $\tau$. In the limit of fast relaxation, $1/\tau \rightarrow \infty$, the distribution $P_n$  approaches the equilibrium distribution $P_n^{\mrm{eq}}$, while in the opposite limit of slow relaxation, $1/\tau \rightarrow 0$, the phonon distribution function is entirely controlled by electron-induced processes. 
 
For small inter-dot tunneling, we can evaluate the transition rates $W^{n \rightarrow n^{\prime}}$ by Fermi's Golden Rule, working to lowest order in the hopping amplitude $t_0$. By accounting for tunneling processes between the quantum dots going in both directions, the rates can be expressed as $W^{n \rightarrow n^{\prime}}=\sum_{\alpha \neq \beta} W_{\alpha \beta}^{n \rightarrow n^{\prime}}$ with  
\begin{align}
W_{\alpha \beta}^{n \rightarrow n^{\prime}} =\vert M_{n \rightarrow n^{\prime}} \vert^2 \vert t_0\vert^2 I_{\alpha \beta}^{n\rightarrow n^{\prime}}.\label{eq:Wnnprime}
\end{align}
Here, we label the leads by Greek indices, $\alpha = L, R$. The transition rates involve the Franck-Condon matrix elements $M_{n \rightarrow n' }=\langle n' | e^{- \lambda \hat{x}} |n \rangle $.\cite{koch-2006} An explicit evaluation of these matrix elements yields
\begin{align}
  \vert M_{n\rightarrow n^{\prime} } \vert^2 = \mathrm{e}^{-\lambda^2} \left[\lambda^{Q-q}\, \sqrt{q!/Q!}\, L_q^{Q-q}(\lambda^2)\right]^2, \label{eq:Mnnprime}
\end{align}
with the abbreviations $q=\mathrm{min}(n,n')$ and $Q=\mathrm{max}(n,n')$, while $L_m^n(x)$ denotes the generalized Laguerre polynomials.

The electronic contribution to the transition rates in Eq.\ (\ref{eq:Wnnprime}) is
\begin{align}
 &I_{\alpha \beta}^{n\rightarrow n\pm m}= 2 \pi \int \mrm{d} \epsilon f_\alpha(\epsilon) \left[1-f_\beta(\epsilon^\mp)\right] \rho_\alpha(\epsilon)   \rho_\beta (\epsilon^\mp)\label{eq:auxint01},
\end{align}
with $\epsilon^\mp =\epsilon \mp m \omega$. Here, $f_\alpha(\epsilon)=1/\left(\mathrm{e}^{(\epsilon-\mu_\alpha)/T}+1 \right)$ is the Fermi distribution function for lead $\alpha$ (with chemical potential $\mu_{\alpha}$ and temperature $T$). Accounting for the coupling to the leads, the local density of states of the quantum dot $\alpha$ is given by
\begin{align}
\rho_{\alpha}(\epsilon)= \frac{\Gamma_{\alpha}}{2 \pi [(\epsilon- \epsilon_{\alpha})^2 + (\Gamma_{\alpha}/2)^2]},\label{eq:cooling_rhoalpha}
\end{align}
where $\epsilon_\alpha$ is the renormalized level energy and $\Gamma_\alpha$ denotes the lead-induced broadening of the level. 

\begin{figure}[t]
    \includegraphics[width=8.5cm]{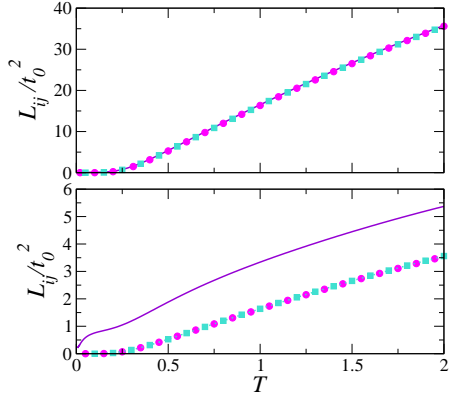}	
\caption{(Color online) Thermoelectric response coefficients $L_{11}$ (solid line), $L_{12}=L_{21}$ (circles), and $L_{22}$ (squares) as function of temperature for weak electron phonon interaction $\lambda=0.1$. The upper (lower) panel corresponds to a coupling to the electron reservoirs of $\Gamma =0.01$ ($\Gamma=0.1$). The quantum-dot levels are in the resonance configuration $\epsilon_R-\epsilon_L = \omega$ with $\epsilon_L=0.025$ and $\epsilon_R=1.025$ with chemical potential $\mu=0$. Energies and temperatures are expressed in units of $\omega$. Differences in units between the response coefficients are also compensated by factors of $\omega$.}
\label{fig:lij}
\end{figure}

\subsection{Thermoelectric response matrix ${\bf L}$}

We can use the rate equations (\ref{eq:rateequation}) to compute the thermoelectric linear-response matrix ${\bf L}$. Within the rate-equation formalism, the charge current between the electron reservoirs and the heat current out of the phonon bath can be expressed as
\ba
J^C & = & e \sum_{n,n^{\prime}} \left( P_n^{\rm eq} W^{n \rightarrow n^{\prime}}_{L R} -  P_{n^{\prime}}^{\rm eq} W^{n^{\prime} \rightarrow n}_{R  L} \right),  \\
J^Q & = & \frac{\omega}{2}  \sum_{n,n^{\prime}} (n-n^{\prime}) \left( P_n^{\rm eq} W^{n \rightarrow n^{\prime}} -  P_{n^{\prime}}^{\rm eq} W^{n^{\prime} \rightarrow n} \right). \nonumber
\ea
In linear response, the phonon distribution function is close to equilibrium (at temperature $T_{\rm ph}$) at all times. Hence, these expressions involve the equilibrium distribution function $P_n^{\rm eq}$ given in Eq.\ (\ref{peq}). One readily verifies that in strict equilibrium, i.e., for $eV= \Delta T = 0$, the rates satisfy detailed balance, $P_n^{\rm eq} W^{n\rightarrow n^{\prime} }_{\alpha  \beta} = P_{n^{\prime} }^{\rm eq} W^{n^{\prime} \rightarrow n}_{\beta  \alpha} $, implying $J^C=J^Q=0$ as expected.

\begin{figure}[tb]
  \begin{center}
        \includegraphics[width=7cm]{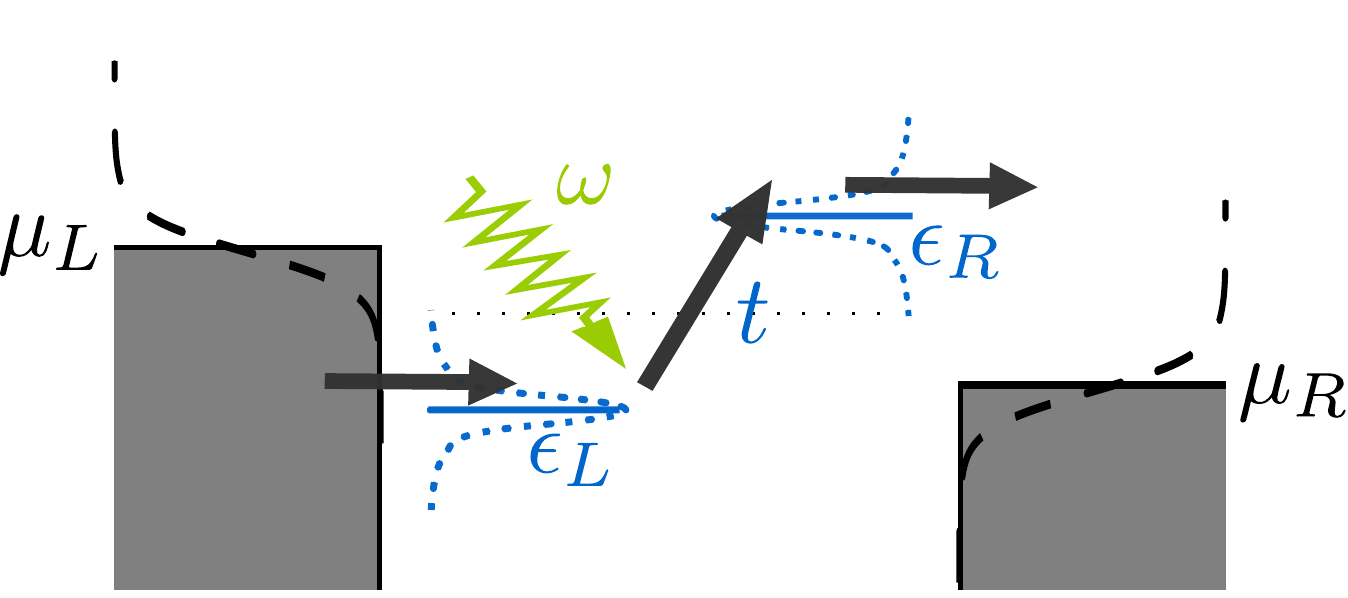}
  \end{center}
\caption{(Color online) Sketch of the phonon-absorption process in the resonant situation $\epsilon_R=\epsilon_L+\omega$.}
\label{fig:absorption}
\end{figure}

We can now work out the charge and heat currents to linear order in $eV$ and $\Delta T$. After some algebra, this yields  
\ba \label{lij}
L_{11} & = & (2 \pi) |t_0|^2 (1-e^{-\beta \omega}) \sum_{n,n^{\prime}}  \lambda_{n,n^{\prime}}, \nonumber \\
L_{12} &=&  (2 \pi) \omega  |t_0|^2 (1-e^{-\beta \omega}) \sum_{n,n^{\prime}} (n-n^{\prime})  \lambda_{n,n^{\prime}},
 \\
L_{22} &=& (2 \pi)  \omega^2 |t_0|^2 (1-e^{-\beta \omega}) \sum_{n,n^{\prime}} (n-n^{\prime})^2   \lambda^{s}_{n,n^{\prime}}, \nonumber
\ea
for the thermoelectric response functions. Here, we used the shorthands
\ba
\lambda_{n,n^{\prime}} &=&  |M_{n \rightarrow n^{\prime}} |^2  \int_{-\infty}^{\infty} d \epsilon  e^{\beta(\epsilon -\mu)} F_{n,n^{\prime}}(\epsilon) 
\rho^{n,n^{\prime}}_{LR}(\epsilon),
\nonumber \\
\lambda^s_{n,n^{\prime}} &=&  \frac{1}{2} |M_{n \rightarrow n^{\prime}}|^2  \int_{-\infty}^{\infty} d \epsilon  e^{\beta(\epsilon -\mu)} F_{n,n^{\prime}}(\epsilon) \nonumber \\
& &\times  \left(\rho^{n,n^{\prime}}_{LR}(\epsilon)+ \rho^{n,n^{\prime}}_{RL}(\epsilon) \right)
\ea
with
\ba
 F_{n,n^{\prime}}(\epsilon) & = & f(\epsilon+n\omega) f(\epsilon+n^{\prime}\omega), \nonumber \\
 \rho^{n,n^{\prime}}_{\alpha \beta}(\epsilon) & = & \rho_{\alpha}(\epsilon+n^{\prime}\omega) \rho_{\beta}(\epsilon+n \omega),
\ea
in terms of $f(\epsilon)=1/[e^{\beta(\epsilon-\mu)}+1]$. It can be verified that these expressions satisfy the Onsager relation $L_{12}=L_{21}$. 

To illustrate the response functions $L_{ij}$, we calculate them explicitly as functions of the electron temperature for a system in the resonant configuration $\epsilon_R-\epsilon_L = \omega$ and with a chemical potential which is located slightly below the energy of the left quantum-dot level. We also express all energies in units of $\omega$. The results are shown in Fig.\ \ref{fig:lij}, for the case of a weak electron-phonon coupling  $\lambda$. The two panels correspond to different values of the coupling between the dots and the reservoirs $\Gamma$. For this resonant configuration of the quantum dot levels, the dominant tunneling process involves the absorption of one phonon quantum, as sketched in Fig.\ \ref{fig:absorption}. Thus, the sums in Eq.\ (\ref{lij}) are dominated by $n=1$ and $n^{\prime}=0$ for sufficiently weak $\Gamma$. Hence, $L_{22}\sim \omega L_{12} \sim \omega^2 L_{11}$ (see upper panel).  As $\Gamma$ increases (see lower panel of the figure), direct tunneling without the absorption of a phonon quantum becomes more likely, in addition to the tunneling with phonon absorption. The direct tunneling process corresponds to terms with $n=n^{\prime}=0$ in Eq.\ (\ref{lij}) and thus contributes only to $L_{11}$. Hence, for larger hybridizations $\Gamma$, the thermoelectric coefficients satisfy $L_{11} > \omega L_{12}$ and $L_{12} \sim \omega L_{22}$.

\begin{figure}[t]
    \includegraphics[width=8.5cm]{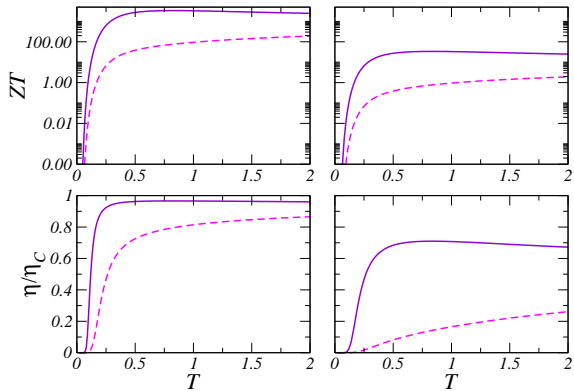}	
\caption{(Color online) Figure of merit $ZT$ and efficiency $\eta$ of the refrigerator for different hybridization strengths $\Gamma_L=\Gamma_R=\Gamma =0.01,0.1$ (left and right panels, respectively). Solid (dashed) lines correspond to electron-phonon interaction $\lambda=1$ and $\lambda=0.1$, respectively. Other parameters are as in Fig.\ \ref{fig:lij}.}
\label{fig:eta-q}
\end{figure}

\subsection{Efficiency and figure of merit}

Based on the coefficients $L_{ij}$, we can now calculate the efficiency with which heat can be extracted from the phonon bath [see Eq.\ (\ref{eta})] as well as the associated figure of merit [Eq.\ (\ref{z})]. As we will see, the result of the previous section that the ratios of the response coefficients are of order unity (when made dimensionless by appropriate powers of $\omega$) implies both a large figure of merit and efficiencies which are near the Carnot limit. 

Representative results for the efficiency and the figure of merit are presented in Fig.\ \ref{fig:eta-q}, for the same resonant configuration discussed in the previous section. We observe that large values of $ZT$ and efficiencies close to the Carnot limit are attained for small hybridizations between quantum dots and leads (see left panels of Fig.\ \ref{fig:eta-q}). Then, the dominant process is the one sketched in Fig.\ \ref{fig:absorption}, which corresponds to electron tunneling accompanied by single-phonon absorption. The strength of this process increases with increasing electron-phonon interaction $\lambda$. Hence, the efficiency increases with $\lambda$. This can be deduced by comparing the solid and the dashed lines in the plots, which correspond to higher and lower $\lambda$, respectively. In contrast, increasing the hybridization with the reservoirs characterized by $\Gamma$,  tends to decrease the efficiency (compare left and right plots), as increasing $\Gamma$ increases the direct tunneling of electrons without phonon absorption.

\subsection{Effective temperature}

We now turn to characterize the refrigeration device by the effective (nonequilibrium) temperature of the phonon mode. This characterization is appropriate when $\tau \to \infty$, i.e., when the heating of the phonon mode by the coupling to the phononic bath becomes negligible relative to the electron-induced processes. 

\begin{figure}[t]
	\includegraphics[width=9cm]{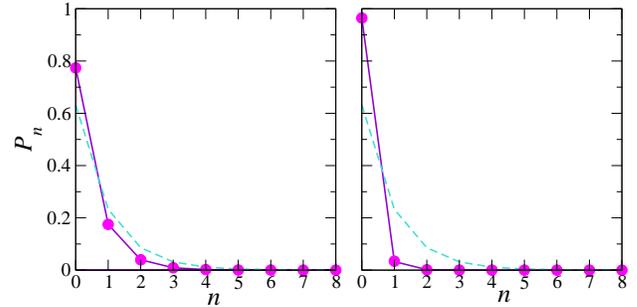}
\caption{(Color online) Points connected with solid lines: Probability distribution function $P_n$ calculated from the numerical solution of Eq.\ (\ref{eq:rateequation}) in the stationary non-equilibrium case by considering up to $13$ phonon modes. Circles: Equilibirum
probability distribution function $P_n^{\rm eq}$ [see Eq.\ (\ref{peq})] corresponding to a phonon temperature $T_{\rm ph}=T^{\rm eff}$, where the effective temperature is defined in Eq.\ (\ref{eq:Teff}).
Dashed lines: $P_n^{\rm eq}$ for $T_{\rm ph}= T$. Left (right) panel corresponds to $T=1$, $V=0.5$  ($T=1$, $V=2.5$) in which case $T_{\rm eff}/T=0.67$ ($T_{\rm eff}/T=0.3$). We are considering a small relaxation rate $1/\tau$ ($\tau \vert t_0\vert^2 \lambda^2/\omega = 10$),  $t_0=0.1$, and weak coupling to the reservoirs,  $\Gamma=0.01$. All other parameters are as in Fig.\ \ref{fig:lij} and all the energies are expressed in units of $\omega$.}
\label{fig:pn}
\end{figure}

\begin{figure*}[tb]
  \begin{center}
\includegraphics[width=6cm]{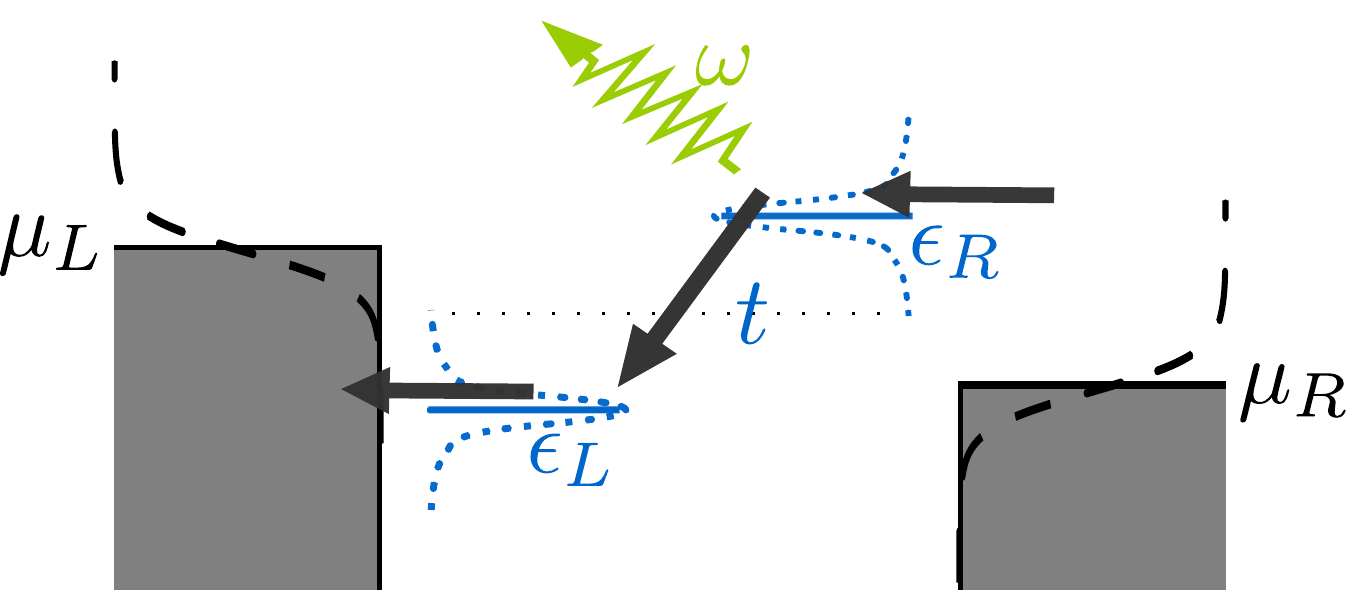}
\includegraphics[width=6cm]{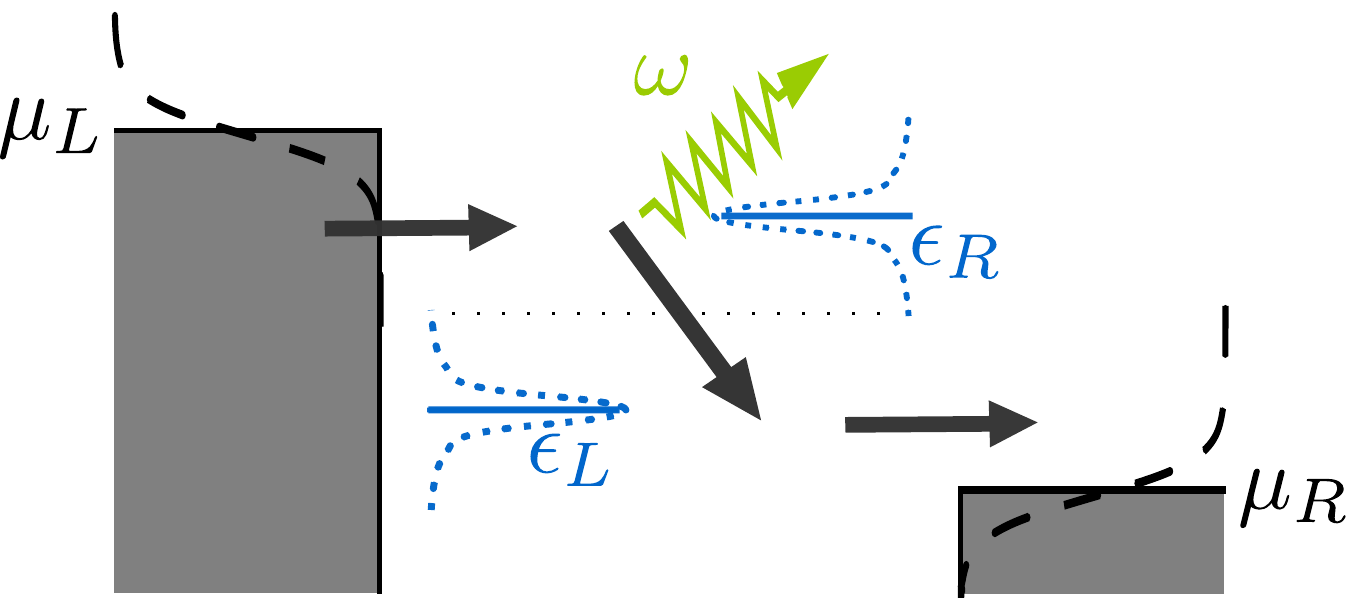}
  \end{center}
\caption{(Color online) Phonon-emission processes which limit the phonon refrigeration in the resonant situation $\epsilon_R=\epsilon_L+\omega$. 
The process on the left (right) is dominant for $eV< \omega$ ($eV>\omega$).}
\label{fig:emission}
\end{figure*}

\subsubsection{Linear response}

In linear response, we can use Eq.\ (\ref{tloc}) for the effective temperature. Thus, the effective temperature depends on the ratio $L_{12}/L_{22}$. As discussed in the previous section and shown in Fig.\ \ref{fig:lij}, this ratio approximately equals $1/\omega$ in the resonant configuration $\epsilon_R-\epsilon_L=\omega$  when the single phonon absorption process dominates and it is basically independent of 
the hybridization $\Gamma$ and the electron-phonon coupling $\lambda$.  Thus, we have
\be \label{tlocsmall}
  T^{\rm eff} \simeq T \left(1-eV/\omega \right),
\ee
over a wide range of parameters in the linear-response regime ($eV\ll\omega$). 

\subsubsection{Beyond linear response}

Beyond linear response, we can characterize the cooling power by directly computing the effective temperature from the rate equations. Here, it is more convenient to define the effective temperature based on the stationary (nonequilibrium) phonon distribution $P^{\mrm{stat}}_n$ through
\begin{align}
 T^{\rm eff} = \frac{\omega}{ \ln(1+1/\overline{n}^{\mrm{stat}})}, \label{eq:Teff}
\end{align}
where $\overline{n}^{\mrm{stat}} = \sum_n n P^{\mrm{stat}}_n$ is the average phonon excitation corresponding to the stationary phonon distribution. Thus, Eq.\ (\ref{eq:Teff}) simply mimics the relation between temperature and average phonon excitation for a thermal equilibrium distribution like Eq.\ (\ref{peq}) with $T_{\rm ph}= T^{\rm eff}$. Notably, in the linear-response regime this definition of the effective temperature coincides with the previous definition based on the heat current from a fictitious phonon reservoir (a thermometer).

We note in passing that similar definitions were used, e.g., in Refs.\ \onlinecite{pistolesi-2009,leto1,leto2,hugo,lili,eftem,galptef}. One can readily check the usefulness of this definition of effective temperature {\em a posteriori} by comparing the full stationary phonon distribution with the thermal equilibrium distribution for temperature $T^{\rm eff}$. An example is shown in Fig.\ \ref{fig:pn}, where the exact
non-equilibrium distribution function, obtained by the numerical solution of  Eq.\ (\ref{eq:rateequation}) in the stationary case (see solid lines) is shown along with the thermal distribution function $P^{\rm eq}_n$ of Eq.\ (\ref{peq}), corresponding to $T_{\rm ph}= T^{\rm eff}$ (circles). The  thermal distribution function corresponding to $T_{\rm ph}=T$ is also shown for comparison (see dashed lines). 

An important question which can be addressed with this definition concerns the processes which limit the cooling power of the device and hence the 
limiting temperature which can be reached. Considering again the resonant configuration discussed above, we can make analytical progress in the limit of weak electron-phonon coupling $\lambda$ where  single-phonon processes dominate. In this regime the relevant rates involved are
\begin{align}
  W^{n \rightarrow n\pm1} \simeq \mathrm{max}(n,n\pm1) \, \lambda^2 \vert t_0\vert^2 I^{\pm},
\end{align}
where we abbreviate $I^{\pm}=I^{n \rightarrow n\pm1}$. This allows one to rewrite the rate equation \eqref{eq:rateequation} in terms of the average phonon number, $\overline{n} =\sum_n n P_n$, which yields
\begin{align}
  \dot{\overline{n}} \simeq \left[\lambda^2 \vert t_0\vert^2(I^+ - I^-) - 1/\tau \right] \overline{n} + \lambda^2 \vert t_0\vert^2 I^+ + \overline{n}^{\mrm{eq}}/\tau.
\end{align}
Here, $\overline{n}^{\mrm{eq}}=\left(\mrm{e}^{\beta \omega}-1\right)^{-1}$ denotes the average phonon number in equilibrium at the bath temperature. This equation readily yields the stationary solution 
\begin{align}\label{eq:nstat01}
  \overline{n}^{\mrm{stat}} &= \overline{n}_0^{\mrm{stat}} + \frac{1}{\lambda^2 \vert t_0 \vert^2 \tau} \frac{\overline{n}^{\mrm{eq}}-\overline{n}_0^{\mrm{stat}}}{I^- -I^+ + 1/(\lambda^2 \vert t_0 \vert^2 \tau)},
\end{align}
where $\overline{n}_0^{\mrm{stat}} = I^+/(I^- - I^+)$ denotes the solution in the absence of a phonon bath ($\tau\to\infty$).  
In order to cool the system, \textit{i.e.}, \ $\overline{n}^{\mrm{stat}} < \overline{n}^{\mrm{eq}}$, it is necessary that $\overline{n}_0^{\rm stat} < \overline{n}^{\mrm{eq}}$. Thus, this condition will allow us to identify the temperature range for which cooling is possible.

The integrals $I^{\pm}_{\alpha \beta}$ become particularly simple when assuming the limit $\Gamma \ll T$ which allows one to obtain analytical results (see App.\ \ref{appint} for details). The leading absorption process is depicted in Fig.\ \ref{fig:absorption}. The important emission processes are shown in Fig.\ \ref{fig:emission}, with their relative magnitude depending on the ratio of voltage and phonon frequency.

For small voltages, $eV < \omega$, the phonon absorption mainly competes with the phonon emission process in upstream tunneling of electrons between the dots, see Fig.\ \ref{fig:emission} (left). This process is possible  at finite temperature (though exponentially suppressed) due to the thermal broadening of the Fermi functions of the leads. Specifically, for $1/\tau \rightarrow 0$, one has $I^- \gg I^+$, so that 
$\overline{n}^{\mathrm{stat}} \simeq I^+/I^-$ and we find an effective temperature 
\begin{align}
 \frac{ T^{\rm eff}}{T} \simeq \frac{\omega/T}{\ln(1+\e^{(\omega+eV)/T})}\simeq \frac{\omega}{\omega+eV}.\label{eq:Teff_approx01}
\end{align}
Thus, the transport current does indeed cool the phonon mode, with the effective temperature decreasing for increasing bias voltage. It is interesting to note that this result reduces to the earlier linear-response result Eq.\ (\ref{tlocsmall}) when $eV \ll \omega$.

For larger biases, $eV >\omega$, there is an additional heating channel associated with downstream tunneling of electrons, see Fig.\ \ref{fig:emission} (right). This process actually dominates heating when $T \ll eV$. Consequently, we find 
\begin{align}
  T^{\rm eff} \simeq \frac{\omega}{2\ln(2\omega/\Gamma)}, \label{eq:Teff_approx02}
\end{align}
This shows that for $eV > \omega$, cooling is possible only for relatively large temperatures $\omega/[2\ln(2\omega/\Gamma)] <T< eV $.

\subsubsection{Beyond linear response -- numerical results}

\begin{figure}[t]
  \begin{center}
    \includegraphics[width=7cm]{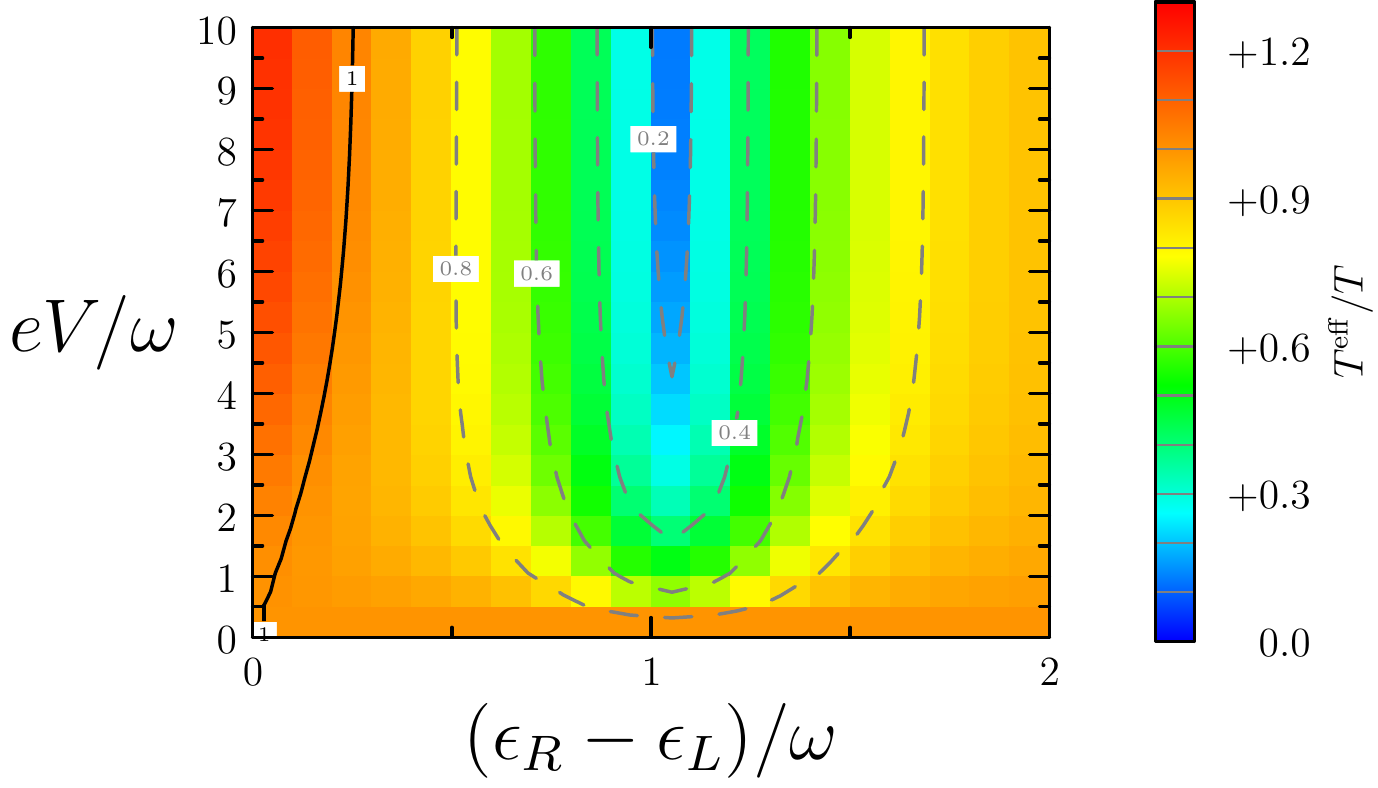}
  \end{center}
\caption{(Color online) $T^{\rm eff}/T$ {\em vs} detuning of the electronic energy levels and bias voltage, for weak electron-phonon coupling $\lambda=0.1$, small relaxation rate $1/\tau$ ($\tau \vert t_0 \vert^2 \lambda^2/\omega = 10$), $\Gamma/\omega=0.01$,  and $T/\omega=1$. }
\label{fig:Teff_eReL}
\end{figure}

We complete this section with numerical results for the effective temperature $T^{\rm eff}$ which cover a wider range of parameters than accessible analytically. The efficiency of the refrigerator depends crucially on the electronic levels. This is illustrated in Fig.\ \ref{fig:Teff_eReL} for a small relaxation rate $1/\tau$. One observes that the energy-level configuration which is most favorable for cooling is indeed the resonant configuration $\epsilon_R -\epsilon_L=\omega$ sketched in Fig.\ \ref{fig:absorption} (with $\mu_L=\mu_R+eV$). More generally, we find that cooling is possible for a wide range of values of the ratio $(\epsilon_R-\epsilon_L)/\omega$ as well as a wide range of voltages $V$. We verified that the lowest effective temperatures are achieved at resonance and high voltages $eV$, in agreement with Eqs.\ (\ref{eq:Teff_approx01}) and (\ref{eq:Teff_approx02}).  Increasing the phonon relaxation rate $1/\tau$ or the degree of coupling of the quantum dots to the electron reservoirs favor the thermalization with the external bath. Thus, these effects work against cooling as can be easily verified numerically. Changing the electron-phonon coupling $\lambda$ does not introduce relevant qualitative changes in the behavior of $T^{\rm eff}$ over a wide range of values.

\section{Classical regime} 
\label{sec:classical_regime}

\subsection{Langevin dynamics and thermoelectric response functions}

We now turn to the classical regime in which the phonon frequency is small compared to the rate at which electrons are passing between the reservoirs, $\omega\ll\Gamma$ and $\omega \ll t_0$. In this regime, the vibrational dynamics can be described in terms of a Langevin equation 
\begin{equation}
  M \ddot X  = - M \omega^2 X + F(X) -\gamma \dot{X} +\xi(t) \,,
   \label{langevin} 
\end{equation}
where $M$ is the effective mass of the vibrational mode, which is now represented by the classical coordinate $X(t)$. The first term on the right-hand side is the elastic restoring force of the vibrational mode. The remaining terms originate from coupling to the electrons and the phonon reservoir. The Born-Oppenheimer force $F(X)$ can be accounted for by a renormalization of the elastic restoring force which will be left implicit in the following. 

Both the friction coefficient $\gamma$ and the stochastic force $\xi(t)$ have (additive) contributions from the coupling to the electrons and the phonon reservoir,
\ba
\gamma & = & \gamma_{\rm e} + \gamma_{\rm ph}, \nonumber \\
\xi(t) & = & \xi_{\rm e}(t) + \xi_{\rm ph}(t),
\ea
When a phonon reservoir is present, it by definition keeps the vibrational mode in thermal equilibrium at all times. Then, the phonon contributions $\gamma_{\rm ph}$ and $\xi_{\rm ph}$ are much larger than their electronic counterparts. The fluctuating forces are characterized by the correlators 
\ba
\langle \xi_{\rm e} (t)  \xi_{\rm e} (t)  \rangle  & = & D_{\rm e } \delta( t- t^{\prime}),  \nonumber \\
\langle \xi_{\rm ph} (t)  \xi_{\rm ph} (t)  \rangle & = &  D_{\rm ph } \delta( t- t^{\prime}), \nonumber \\
 \langle \xi_{\rm e} (t)  \xi_{\rm ph} (t)  \rangle &  = &  0.
\ea
The phonon reservoir is in thermal equilibrium at the phonon temperature $T_{\rm ph}$. Thus, the fluctuation-dissipation theorem implies that 
\begin{equation}
  D_{\rm ph} = 2\gamma_{\rm ph} T_{\rm ph}.
\end{equation}
At the same time, when a voltage bias is applied to the electronic system, the corresponding friction and fluctuation coefficients contain a nonequilibrium contribution (labelled by the superscript ne) in addition to the equilibrium one (labelled by eq),
\begin{eqnarray} \label{cofe}
  \gamma_{\rm e} &=& \gamma_{\rm e}^{\rm eq} + \gamma_{\rm e}^{\rm ne} \\
  D_{\rm e} &=& D_{\rm e}^{\rm eq} + D_{\rm e}^{\rm ne}.
\end{eqnarray}
Again, the fluctuation-dissipation theorem implies the relation 
\begin{equation}
  D^{\rm eq}_{\rm e} = 2\gamma^{\rm eq}_{\rm e} T
\end{equation}
in terms of the electron temperature $T$. 

The coefficients $\gamma_{\rm e}$ and $D_{\rm e}$ can be evaluated for the microscopic model under consideration. Before doing so, we present a general derivation of the thermoelectric response functions from the Langevin dynamics. We start by considering the heat current $J^Q$ flowing from the phonon  to the electron reservoir. This heat current is effected by the coupling between vibrational mode and electrons as encoded in the friction $\gamma_{\rm e}$ and the force correlator $D_{\rm e}$. We can compute the heat current by multiplying the Langevin equation by $\dot X$ and rewriting it as an equation for the time derivative of the energy stored in the vibrational mode. In a stationary state, this energy is time independent on average, with the heat current lost to the electron system compensated by the phonon reservoir. Thus,
\begin{eqnarray}
  J^Q &=& -\left[ \frac{d}{dt} \langle \frac{1}{2} M \dot{X}^2+ \frac{1}{2} M \omega^2 X^2 \rangle \right]_{\rm e}  \nonumber \\
      &=&   \gamma_{\rm e} \langle \dot{X}(t)^2 \rangle - \langle \xi_{\rm e}(t) \dot{X}(t) \rangle.
\end{eqnarray}
As we are assuming that the vibrational mode is fully thermalized with the phonon reservoir, 
by equipartition  we have $\langle \dot{X}(t)^2 \rangle = T_{\rm ph}/M$. The fluctuating contribution $\delta X_{\rm e}(t)$ of the vibrational coordinate is 
 proportional to $\xi_{\rm e}(t)$. Hence
\begin{equation}
  \delta X_{\rm e}(\Omega) =  \frac{1}{- M\Omega^2 + M\omega^2 +i\gamma\Omega} \xi_{\rm e}(\Omega).
\end{equation}
Here, the last relation follows from the Langevin equation written in Fourier space. Inserting this into the correlator $\langle \xi_{\rm e}(t)\dot{X}(t)\rangle$, using the correlator of the fluctuating force $\xi_{\rm e}$, and performing the frequency integration, we obtain
\begin{equation}
   J^Q = \frac{\gamma_e T_{\rm ph}}{M} - \frac{D_{\rm e}}{2M}.
\end{equation}
We can now use this general expression to compute both $L_{21}$ and $L_{22}$.  

To compute $L_{22}$, we first assume a small temperature difference $\Delta T = T_{\rm ph}-T$ between phononic and electronic reservoirs, but zero applied bias $V$. Then, the electrons are in thermal equilibrium and we can use the fluctuation-dissipation theorem to replace $D_{\rm e}$. This yields \begin{equation}
   J^Q = \frac{\gamma_e^{\rm eq}}{M}(T_{\rm ph} - T)
\end{equation}
and thus
\begin{equation}
   L_{22} = \frac{\gamma_e^{\rm eq} T^2}{M}.
   \label{l22slow}
\end{equation}
Similarly, we consider $T_{\rm ph} = T$ with nonzero applied bias $V$ to compute $L_{21}$. Then, the equilibrium contributions of $\gamma_{\rm e}$ and $D_{\rm e}$ cancel in $J^Q$ and only the nonequilibrium contributions remain,
\begin{equation}
   J^Q = \frac{\gamma_e^{\rm ne} T}{M} - \frac{D^{\rm ne}_{\rm e}}{2M}.
\end{equation}
This yields
\begin{equation}
   L_{21} = \frac{T}{2M} \left[\frac{d}{d(eV)}(2\gamma_{\rm e}^{\rm ne} T - D^{\rm ne}_{\rm e})\right]_{eV=0}.  
\label{l21slow}
\end{equation} 
Note that by Onsager's relation, $L_{12}=L_{21}$. Finally, we remark that up to factors of temperature, $L_{11}$ is simply the conductance of the system.  

\subsection{Results for the microscopic model}

We are now in a position to derive expressions for the thermoelectric response coefficients as functions of the parameters of the microscopic model. As shown in previous works, both the friction coefficient and the force correlator can be conveniently evaluated in terms of the Green functions of the the underlying microscopic model,\cite{pisto-class} or alternatively in terms of the electronic scattering matrices.\cite{bode-2011,bode-2012} For weak electron-phonon coupling $\lambda$, the interdot hopping amplitude [cf.\ Eq.\ (\ref{interdot})] becomes 
\begin{equation} \label{tx}
t(X) \simeq t_0 (1 - \lambda X),
\end{equation}
and the general expressions yield
\begin{align} \label{coef}
\gamma_{\rm e} (X)=& (t_0\lambda)^2 \sum_{\alpha, \beta} \mbox{Re} \left[\int_{-\infty}^{\infty} \frac{d \epsilon}{2 \pi} \partial_{\epsilon} 
G^{f,>}_{\alpha, \beta} (\epsilon, X) G^{f,<}_{\overline{\beta}, \overline{\alpha}} (\epsilon, X) \right], \nonumber \\
D_{\rm e} (X)=& (t_0 \lambda)^2 \sum_{\alpha, \beta} \mbox{Re}\left[ \int_{-\infty}^{\infty} \frac{d \epsilon}{2 \pi} 
G^{f,>}_{\alpha, \beta} (\epsilon, X) G^{f,<}_{\overline{\beta}, \overline{\alpha}} (\epsilon, X) \right].
\end{align}
Here, $\alpha, \beta= L, R$ and $\overline{L}=R,\; \overline{R}=L$. We note that in general, these coefficients depend on the vibrational coordinate $X$. In the following, we assume the limit of small oscillations so that we can linearize the Langevin equation about the vibrational equilibrium and approximate $\gamma_e$ and $D_e$ by their values at $X=0$. The integrands involve the lesser and greater Green functions, which are given in App.\ \ref{froz}. We find it convenient to express the results in terms of the partial densities of states 
\be \label{eq:rhoalphabeta}
 \rho_{\alpha \beta}^{\delta}(\epsilon)= G^{f, R}_{\alpha \delta}(\epsilon) \Gamma_{\delta} G^{f, A}_{\delta \beta }(\epsilon).
\ee 
and the associated total density of states
\be \label{rho}
 \rho_{\alpha \beta}(\epsilon)= \sum_{\delta=L,R}\rho_{\alpha \beta}^{\delta}(\epsilon).
\ee
This yields 
\begin{align} \label{coefeq}
\gamma^{\rm eq}_{\rm e} =& - \frac{(t_0 \lambda)^2}{2} \sum_{\alpha, \beta} \int_{-\infty}^{\infty} \frac{d \epsilon}{2 \pi} \partial_{\epsilon} f(\epsilon) 
\rho_{\alpha, \beta} (\epsilon) \rho_{\overline{\beta}, \overline{\alpha}} (\epsilon) , \nonumber \\
D^{\rm eq}_{\rm e} =& (t_0 \lambda)^2 \sum_{\alpha, \beta}  \int_{-\infty}^{\infty} \frac{d \epsilon}{2 \pi} f(\epsilon) \left[ 1- f(\epsilon) \right] \rho_{\alpha, \beta} (\epsilon) \rho_{\overline{\beta}, \overline{\alpha}} (\epsilon),
\end{align}
for the equilibrium contributions and 
\begin{align}
\gamma^{\rm ne}_{\rm e} &= -eV (t_0 \lambda)^2 \sum_{\alpha, \beta} \int_{-\infty}^{\infty} \frac{d \epsilon}{2 \pi} \partial_\epsilon f (\epsilon)\partial_\epsilon \rho_{\alpha \beta}(\epsilon) \rho^L_{\overline{\beta} \overline{\alpha}}(\epsilon)
, \nonumber \\
D^{\rm ne}_{\rm e}&= -T eV (t_0 \lambda)^2 \sum_{\alpha, \beta} \int_{-\infty}^{\infty} \frac{d \epsilon}{2 \pi} \partial_\epsilon f (\epsilon)\partial_\epsilon \left( \rho_{\alpha \beta}(\epsilon) \rho^L_{\overline{\beta} \overline{\alpha}}(\epsilon)\right)
\end{align}
for the nonequilibrium contributions. One readily confirms that the equilibrium contributions obey the fluctuation-dissipation theorem. 

\begin{figure}[tb]
    \includegraphics[width=8.5cm]{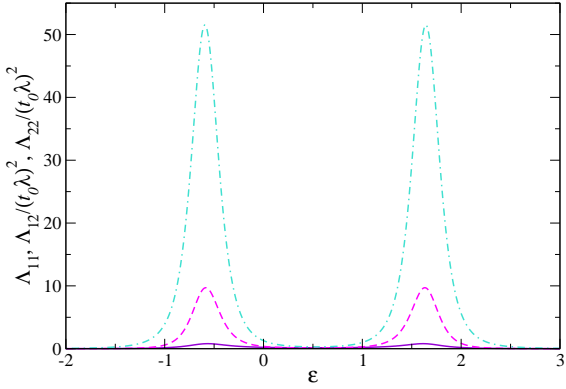}	
\caption{(Color online) $\Lambda_{11}(\epsilon), \Lambda_{12}(\epsilon)/(t_0 \lambda)^2, \Lambda_{22}(\epsilon)/(t_0 \lambda)^2$ defined in Eq.\ (\ref{lambdas}), corresponding, respectively, to solid, dashed and dot-dashed lines,  for  
$\Gamma_L=\Gamma_R=0.5$. The amplitude of the inter-dot hopping is $t_0=1$ and sets the energy scale.
The energies of the levels of the dots are $\epsilon_L=0.025 $ and $\epsilon_R=1.025 $. }
\label{fig:lambda}
\end{figure}
 
Combining these relations with our expressions (\ref{l22slow}) and (\ref{l21slow}) for the thermoelectric response coefficients, we obtain 
\ba \label{lijc}
  L_{11} & = & - T \int_{-\infty}^{\infty} \frac{d \epsilon}{2 \pi}  \partial_{\epsilon} f(\epsilon) \Lambda_{11}(\epsilon) \nonumber \\
  L_{22} & = &-   \frac{T^2 }{M}  \int_{-\infty}^{\infty}  \frac{d \epsilon}{2 \pi}  \partial_{\epsilon} f(\epsilon) \Lambda_{22}(\epsilon) \nonumber \\
  L_{21} & = & L_{12} = - \frac{T^2}{2M} \int_{-\infty}^{\infty} \frac{d \epsilon}{2 \pi}  \partial_{\epsilon} f(\epsilon) \Lambda_{12}(\epsilon)
\label{lij_class}
\ea
with
\ba \label{lambdas}
\Lambda_{11}(\epsilon) & = & \frac{1}{2}\sum_{\alpha} \Gamma_{\alpha} \rho_{\alpha,\alpha}^{\overline{\alpha}}(\epsilon)\nonumber \\
\Lambda_{12}(\epsilon) & = &  (t_0 \lambda)^2 \sum_{\alpha, \beta} \left(  
\partial_\epsilon \rho^R_{\alpha, \beta} (\epsilon) \rho_{\overline{ \beta}, \overline{\alpha}}^L(\epsilon)-
\rho^R_{\alpha ,\beta} (\epsilon) \partial_\epsilon \rho_{\overline{ \beta}, \overline{\alpha}}^L(\epsilon)\right) \nonumber \\
\Lambda_{22}(\epsilon) & = &  \frac{(t_0 \lambda)^2}{2}\sum_{\alpha \beta} \rho_{\alpha, \beta}(\epsilon) \rho_{\overline{\beta},\overline{\alpha}}(\epsilon).
\label{lambdaij_class}
\ea
These expressions can be readily evaluated numerically. 

To provide some intuition, we first plot the integrands $\Lambda_{11}(\epsilon)$, $\Lambda_{22}(\epsilon)$, and $\Lambda_{12}(\epsilon)$.
These functions are shown in Fig.\ \ref{fig:lambda} and have a similar qualitative behavior. In particular, they have peaks at the positions of the electronic levels. We note that in the present regime, where we consider a large hopping parameter $t_0$ between the two dots, the energies of the levels of the double-dot structure differ significantly from the bare energies $\epsilon_L$ and $\epsilon_R$. Importantly, these functions are positive, which implies that the direction of the charge and heat currents are fully determined by the temperature and voltage biases $\Delta T$ and $\Delta \mu$, respectively. 

\subsubsection{Efficiency}

We can now evaluate the efficiency $\eta$ and the figure of merit $ZT$ for the refrigerator in the classical regime. We saw above that the efficiency can approach the Carnot limit in the quantum regime. In contrast, the efficiency will typically be far from the Carnot limit in the classical regime. This can be seen by parametric estimates of the response coefficients. Starting with Eqs.\ (\ref{lij_class}) and 
(\ref{lambdaij_class}), we find $L_{11}\sim T$, $L_{22}\sim (\lambda^2 t_0^2 T^2 / M \Gamma^2)$, and $L_{12}\sim \lambda^2 t_0^2 T^3/ M\Gamma^3$. Thus, we obtain
\begin{equation}
  ZT \simeq \frac{L_{12}^2}{L_{11}L_{22}} \sim \left(\frac{\omega}{\Gamma}\right)^2 \left(\frac{t_0}{\Gamma}\right)^2\left(\frac{T\lambda^2}{k}\right),
\label{ZTclass}
\end{equation}
where we used $\omega^2 = k/M$ for the vibrational frequency in terms of the elastic force constant $k$. $ZT$ is small as all three factors on the right-hand side are small. For the first two factors, this follows by the basic relation $\omega \ll t_0 \ll \Gamma$ underlying the classical regime. \cite{note} In addition, the third factor is small whenever we are allowed to linearize the electron-phonon coupling as we did in Eq.\ (\ref{tx}). Indeed, this linearization is allowed when $\lambda X \sim \lambda (T/k)^{1/2} \ll 1$. Note specifically that this implies that we cannot increase $ZT$ arbitrarily by reducing $M$ as one might have naively assumed based on Eq.\ (\ref{lij_class}).

\begin{figure}[t]
    \includegraphics[width=8.cm]{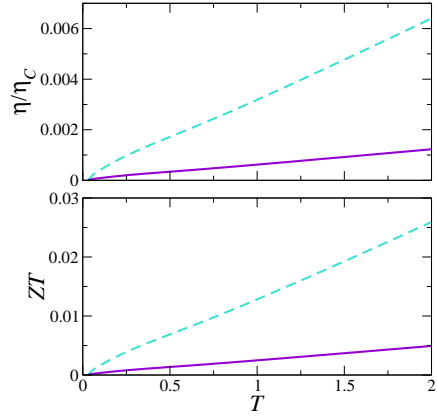}	
\caption{(Color online) Efficiency and figure of merit as a function of the temperature $T$ in the classical regime. We chose $M=1$, $\mu=-0.8$, and 
$\lambda=0.1$.  Solid  and dashed lines correspond to $\Gamma=0.5$ and $\Gamma=1$, respectively.  The unit of energy is set by 
$t_0=1$, while the energies of the levels of the dots are $\epsilon_L=0.025 $ and $\epsilon_R=1.025 $. }
\label{fig:eta_adiabatic}
\end{figure}

These considerations are confirmed by the numerical results shown in Fig.\ \ref{fig:eta_adiabatic} for the figure of merit $ZT$ and the efficiency  $\eta$ in the classical regime. The efficiency is far from the Carnot value, in contrast to the quantum regime, and in agreement with Eq.\ (\ref{ZTclass}), the efficiency and figure of merit increase linearly with temperature. The efficiency also increases with decreasing $\Gamma$. This can be interpreted by noting that the system moves towards the quantum regime when decreasing $\Gamma$. 

\begin{figure}[t]
    \includegraphics[width=9.cm]{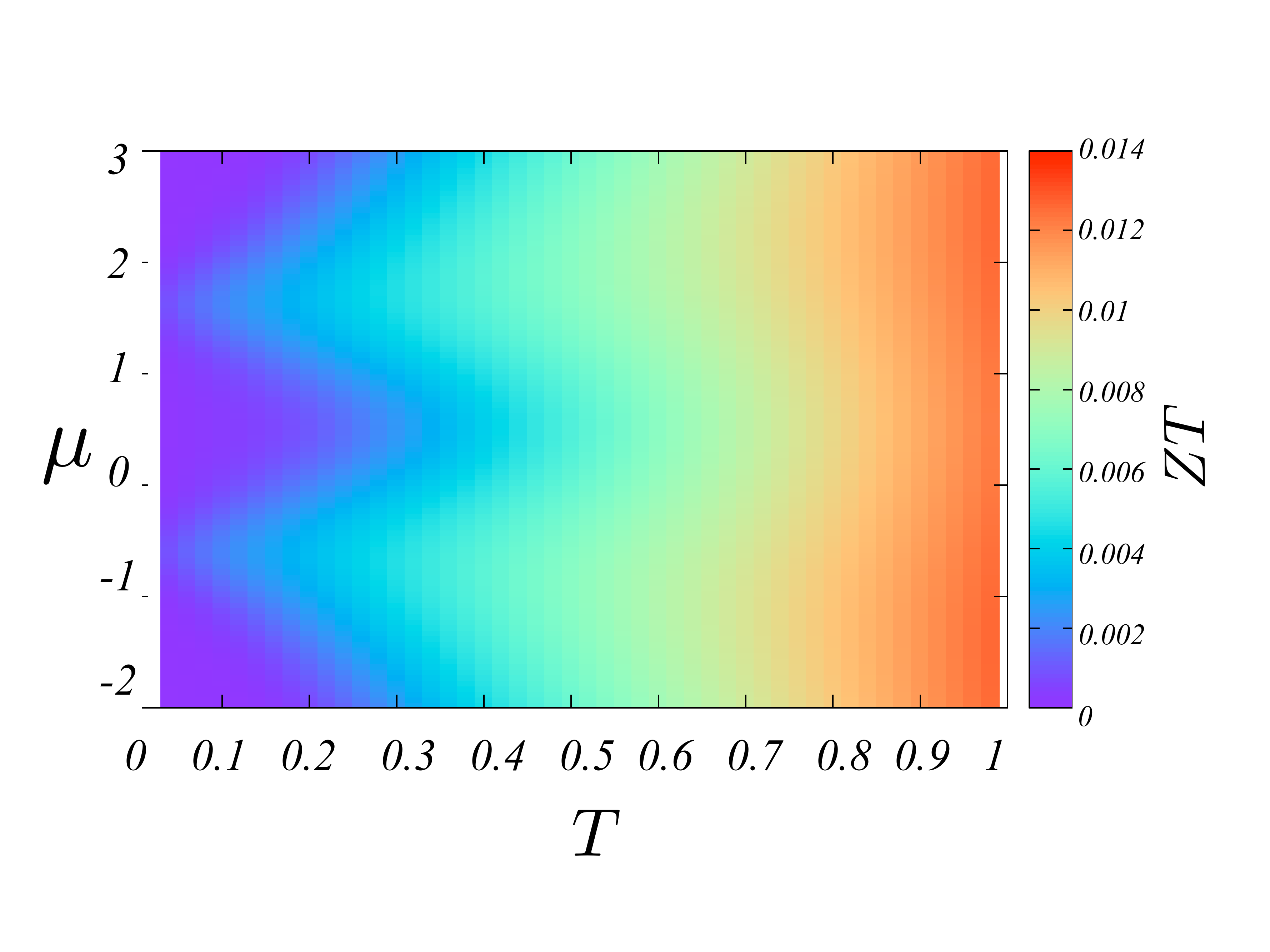}	
\caption{(Color online) Figure of merit {\em vs} temperature $T$ and chemical potential $\mu$ in the classical regime for $\Gamma=0.5$. The mass of the oscillator is $M=1$, the energies of the levels of the dots are $\epsilon_L=0.025 $ and $\epsilon_R=1.025 $, while $t_0=1$ defines the scale for the energies. }
\label{fig:zt_adiabatic}
\end{figure}

In the quantum regime, cooling strongly peaks at the resonance condition $\epsilon_R -\epsilon_L=\omega$. In the classical regime, in contrast, $\omega$ sets the lowest energy scale and is specifically small compared to $\Gamma$ which rules out resonant phonon absorption. Thus, the dependence on  $\epsilon_L$ and $\epsilon_R$ is much weaker. Mainly, the magnitude of  $ZT$ and $\eta$ decrease as $\epsilon_L$ and $\epsilon_R$ approach one another. This just reflects that for $\epsilon_L = \epsilon_R$, the configuration becomes symmetric, which implies $\Lambda_{12}(\epsilon) \rightarrow 0$. Finally, we note that $ZT$ peaks when the chemical potential lies within the peaks of the coefficients $\Lambda_{ij}$ as shown in Fig.\ \ref{fig:zt_adiabatic} (cp.\ Fig.\ \ref{fig:lambda}).  

\subsection{Effective temperature}

Within linear response, the effective temperature in Eq.\ (\ref{tloc}) can be expressed in terms of the thermoelectric coefficients (\ref{lijc}). This yields  
\be \label{teflin}
 T^{\rm eff} \simeq T \left[ 1 +  \frac{eV}{2 \gamma^{\rm eq}} \int d\epsilon \partial_\epsilon f (\epsilon) \Lambda_{12}(\epsilon)\right]. 
\ee
This actually coincides with the effective temperature  
\be \label{etef}
T^{\rm eff}= \frac{ D_ {\rm e}}{ 2 \gamma_{\rm e}}
\ee
motivated by the fluctuation-dissipation relation. Note that here, $\gamma_{\rm e}$ and $D_{\rm e}$ represent non-equilibrium parameters. In fact, expanding the non-equilibrium contributions $D_{\rm e}^{\rm ne}$ and $\gamma_{\rm e}^{\rm ne}$ to linear order in $V$, one recovers Eq.\ (\ref{teflin}) (see Refs.\ \onlinecite{leto1} and \onlinecite{leto2} for a related calculation). Of course, the cooling effect depends on the right direction of current flow. Accordingly, Eq.\ (\ref{teflin}) predicts $T^{\rm eff}<T$ for $eV >0$, and $T^{\rm eff}>T$ for $eV<0$. 

\begin{figure}[t]
    \includegraphics[width=8.5cm]{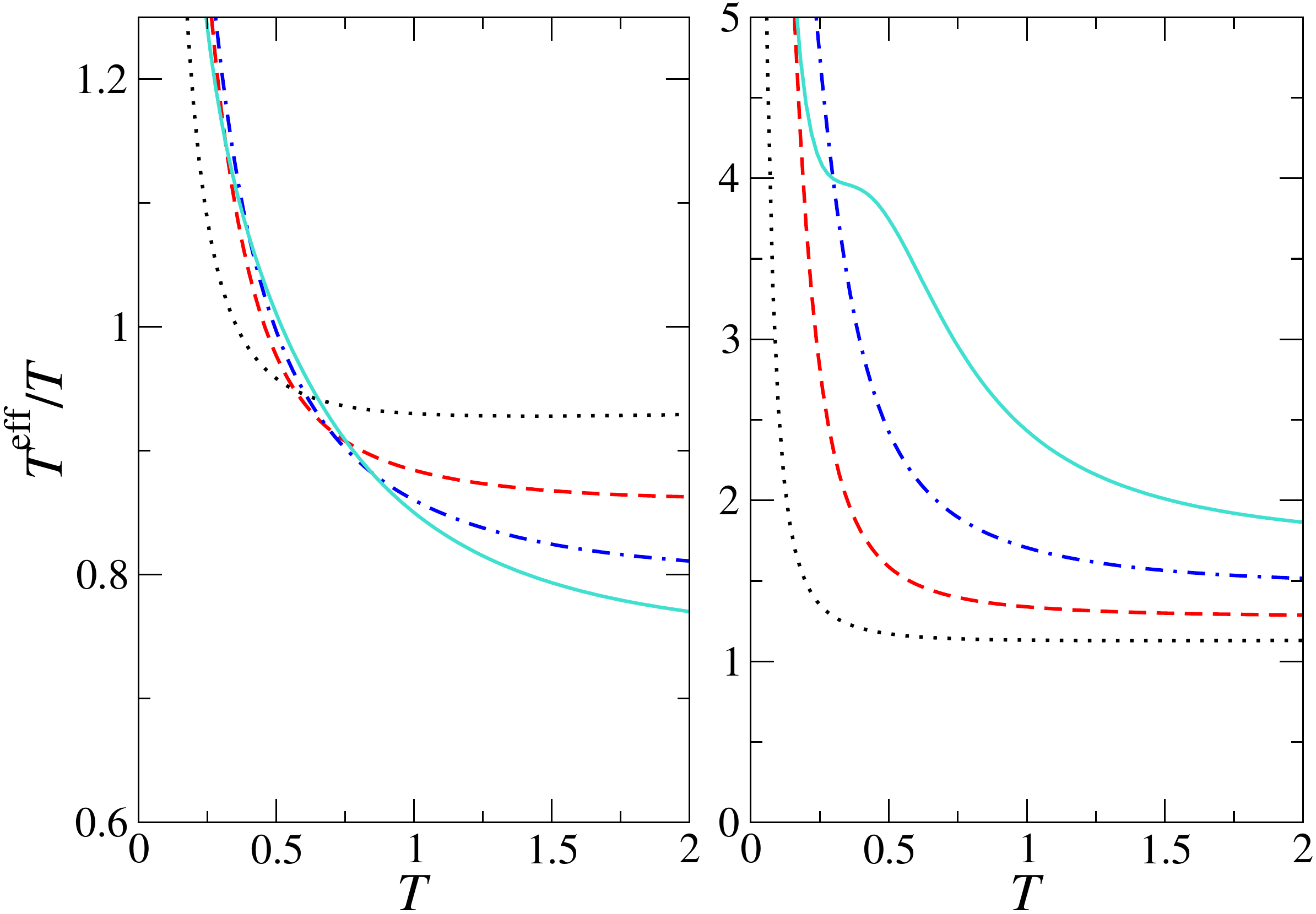}	
\caption{(Color online) $T^{\mrm{eff}}/T$ as a function of the temperature $T$ of the electron reservoirs. Different plots correspond to different voltages $V=0.5,1,1.5,2$. The chemical potentials are $\mu_R=-0.8$ and $\mu_L=\mu_R \pm V$, where the upper (lower) sign corresponds to the  left (right) panel. The energies of the levels of the dots are $\epsilon_L=0.025 $ and $\epsilon_R=1.025 $, while $t_0=1$ defines the scale for the energies as in the previous figures. }
\label{fig:Teff_adiabatic1}
\end{figure}

Numerical results beyond linear response are shown in Fig.\ \ref{fig:Teff_adiabatic1}. The bias voltage is chosen such that the chemical potential is close to the peaks of the function $\Lambda_{ij}(\epsilon)$. In particular, we choose $\mu_R$ slightly below the peak at lower energy and $\mu_L=\mu_R \pm eV$. The left panel corresponds to a level arrangements such that there can be a cooling effect and accordingly, $T^{\rm eff}/T <1$ when the electron temperature $T$ exceeds a threshold value. The right panel of Fig.\ \ref{fig:Teff_adiabatic1} shows results for the opposite configuration $\mu_L=\mu_R - eV$, where cooling is not expected and hnce, $T^{\rm eff}/T \geq 1$ for all $V$ and $T$.

\section{Summary and conclusions} \label{sec:conclusions}

The vibrational motion of a nanoelectromechanical system can be cooled by
a charge current when the electronic levels are appropriately arranged to
favor phonon absorption over emission. In this paper, we have explored
this phenomenon in the context of the general theory of thermoelectrics.
In linear response, this theory allows one to define the efficiency of
cooling as well as the figure of merit $ZT$. We have shown that both concepts
carry over to phonon cooling in nanoelectromechanical systems. As an
important consequence, this allowed us to discuss the efficiency of
these phonon refrigerators in relation to the fundamental Carnot
efficiency.

We have illustrated these concepts for a specific model of a sequential double
quantum dot, arranged such that electron flow is preferentially
accompanied by phonon absorption. Our motivation to study this system was
twofold: First, recent progress in the device fabrication of suspended
carbon nanotubes should make it possible to realize such a structure in
the laboratory. Second, the model is amenable to a largely analytical
treatment with controlled approximations.

We have first considered the limit in which the phonon frequency is large
compared to the rate at which the electrons are passing through the
system. In this {\em quantum limit}, we could describe the electronic and
phononic dynamics by means of master equations. Clearly, in this regime
cooling is most effective when the levels of the two quantum dots are
tuned such that they differ exactly by the phonon frequency, with  increasing level energies in 
the direction of the current flow.
 Indeed, we have shown that in
this case, the efficiency can approach the Carnot efficiency and that the
figure of merit can be very large. Similarly, when the vibrational degree
of freedom is effectively decoupled from a heat bath, the effective
temperature of the phonon mode can be reduced significantly below the
electron temperature.

Second, we have considered the complementary {\em classical regime} in which the
phonons are slow compared to the electron dynamics. Suspended nanotube
devices typically operate in or near this regime when the relevant phonon
mode is the flexural vibrational mode. In this regime, we could describe the
vibrational motion in terms of a Langevin equation which properly accounts
for the nonequilibrium electronic dynamics. We have found that even in this
regime, a double dot structure can operate as a phonon refrigerator but
the typical efficiencies and figures of merit are much reduced compared to
the quantum regime.

Our results not only put recent work on electron-current cooling of vibrational
motion into the context of the general theory of thermoelectrics, but also
provide relevant guidance to future experiments. Most importantly, our
results indicate that efficient cooling of vibrational motion requires an
effort to design structures of suspended carbon nanotube samples closer to the quantum
regime. This could be achieved either by a shorter suspended section or by engineering a larger
string tension. The possibility of efficient cooling  of
vibrational modes with an electron current may also provide interesting applications in refrigerating 
systems into states in which electronic and vibrational degrees of freedom are entangled.

\section{Acknowledgements}
We thank Shahal Ilani and Yuval Oreg for discussions, and acknowledge support from the Alexander von Humboldt Foundation (LA \& FvO), SFB 658 of the Deutsche Forschungsgemeinschft (FvO), as well as CONICET, MINCyT and UBACyT in Argentina (LA).

\appendix

\section{Simplified expressions for the integrals $I^{\pm}_{\alpha \beta}$ in the limit of very small coupling between the dots and the reservoirs. } \label{appint}
In the limit of small coupling between the dots and the reservoirs the the integrals (\ref{eq:auxint01}) with the Lorentzian density of states of Eq.\ (\ref{eq:cooling_rhoalpha})
can be written as
\begin{align}
I^{\pm}_{\alpha \beta} \simeq & F^{\pm}_{\alpha \beta}\frac{2 \Gamma}{(\epsilon_{\beta}-\epsilon_{\alpha} \pm \omega)^2 + (\Gamma/2)^2}  \nonumber \\
& +  \overline{F}^{\pm}_{\alpha \beta} \frac{2 \Gamma}{(\epsilon_{\alpha}-\epsilon_{\beta} \mp \omega)^2 + (\Gamma/2)^2},
\end{align}
with $ F^{\mp}_{\alpha \beta}= f_{\alpha}(\epsilon_{\beta} \pm \omega) [1-f_{\beta}(\epsilon_{\beta})]$ and
$\overline{F}^{\pm}_{\alpha \beta} = f_{\alpha}(\epsilon_{\alpha} ) [1-f_{\beta}(\epsilon_{\alpha} \mp \omega)]$.

We now consider a configuration like the one depicted in Fig.\ \ref{fig:absorption} with the phonon frequency resonant with the separation between the dot levels,
$\omega=\epsilon_R-\epsilon_L$. For small bias voltage, $\mu_L-\mu_R= eV < \omega$,  we assume that only the left level lies within the transport window and we can assume that
it lies perfectly at the center of the transport window. For larger voltages $eV>\omega$, we assume  the two levels are symmetrically aligned with between the two chemical
potentials of the reservoirs.

With these assumptions, it is rather straightforward to get explicit expressions for $I^{\pm}_{\alpha \beta}$ in two limits: (a) The first one corresponds to $T$ setting the smallest energy scale after $\Gamma$, \textit{i.e.} $T\ll \omega$ and $T \ll V$. (b) The other limit corresponds to $T$ setting the largest energy scale, \textit{i.e.} $T \gg \omega$ and $T\gg V$. 

The results for the case (a) are
\begin{align} \label{IRL-approx}
I_{LR}^- & \simeq  \frac{2}{\Gamma},\;\;\;  \forall \; V,\; \omega, \nonumber \\
I_{RL}^- & \simeq  \frac{\Gamma}{(2 \omega)^2} e^{-\beta (V \pm \omega)}, \;\;\;\; {\rm for} \; \;eV <, > \omega, \nonumber \\
I_{LR}^+ & \simeq  \frac{\Gamma}{(2 \omega)^2} e^{-\beta (\omega - V/2)}, \;\;\; {\rm for} \; \; eV < \omega, \nonumber \\
 & \simeq  \frac{\Gamma}{(2 \omega)^2}, \;\;\;{\rm for} \; \; eV > \omega, \nonumber \\
 I_{RL }^+ & \simeq  \frac{2}{\Gamma} e^{-\beta(\omega + V)}, \;\;\; \forall \; V,\; \omega.
\end{align}

The results for the case (b) are
\begin{align} \label{IRL-approx-b}
I_{LR}^- & \simeq   I_{RL}^+ \; \simeq \; \frac{1}{2 \Gamma}, \nonumber \\
I_{RL}^- & \simeq  I_{LR}^+ \; \simeq \; \frac{\Gamma}{8 \omega^2}.
\end{align}

\section{Frozen Green's functions} \label{froz}
The lesser and greater Green's functions entering the forces coefficients are
\be
G^{f, < (>)}_{\alpha,\beta}(\epsilon, X)= \sum_{\gamma} G^{f, R}_{\alpha \gamma}(\epsilon, X) \Sigma^{< (>)}_{\gamma}(\epsilon) G^{f, A}_{\gamma \beta}(\epsilon, X), 
\ee
where  $G^{f, A}_{\alpha \beta}(\epsilon, X)= [G^{f, R}_{\beta \alpha }(\epsilon, X)]^*$ are, respectively, the frozen advanced and retarded Green's functions.  The latter
are
The retarded Green's function becomes (with $\tilde t = t_0 (1 - \lambda X)$)
\begin{equation} \label{gr}
G^{f,R}(\epsilon, X) = \frac{1}{g} \left( \begin{array}{cc} \epsilon-\epsilon_R+\I \Gamma_R &\tilde t\\\tilde t & \epsilon-\epsilon_L+\I \Gamma_L \end{array}\right)\,,
\end{equation}
where $g(X)=(\epsilon-\epsilon_L+\I\Gamma_L)(\epsilon-\epsilon_R+\I\Gamma_R)-\tilde t^2$. The lesser and greater ''self-energies'' are
\be
\Sigma^{< (>)}_{\gamma}(\epsilon)= \lambda^{< (>)}(\epsilon)_{\gamma} \Gamma_{\gamma},
\ee
with $\lambda_{\gamma}^<(\epsilon)= i f_{\gamma}(\epsilon) $ and $\lambda_{\gamma}^>(\epsilon)=- i [1- f_{\gamma}(\epsilon)] $, where $f_{\gamma}(\epsilon)$ is the Fermi function, which depends on the temperature $T$ and chemical potential $\mu_{\gamma}$ of the
reservoir $\gamma$. In what follows, we omit 
explicit reference to the parameter $X$, which enters only in the effective hopping parameter $\tilde{t}$ since we will focus on small $\lambda$, thus $\tilde{t}=t$.

In order to perform an expansion of the different coefficients in $V$, it is useful to recast the lesser and greater functions as follows
\be
G^{f,< (>)}_{\alpha \beta}(\epsilon)= G^{f,{\rm eq},< (>)}_{\alpha \beta}(\epsilon)+  G^{f,{\rm ne},< (>)}_{\alpha \beta}(\epsilon),
\ee
where
\begin{align} \label{glgdes}
 G^{f,{\rm eq},< (>)}_{\alpha \beta}(\epsilon) &= \lambda^{{\rm eq},< (>)}(\epsilon) \rho_{\alpha \beta}(\epsilon),\nonumber \\
  G^{f,{\rm ne},< (>)}_{\alpha \beta}(\epsilon)&= \lambda^{{\rm ne}}(\epsilon) \rho^L_{\alpha \beta}(\epsilon),
 \end{align}
 where the functions $\lambda^{{\rm eq},< (>)}(\epsilon) = \lambda^{< (>)}_{R}(\epsilon)$, which are related to the Fermi distribution function corresponding to the temperature of the reservoirs
 and the chemical potential $\mu_R$, which we take as a reference, while 
 $\lambda^{\rm ne}(\epsilon)  =  i \left[ f_L(\epsilon)- f_R(\epsilon) \right] $, with $\mu_L=\mu_R+eV$. The total and partial densities of states $\rho_{\alpha \beta}(\epsilon)$
 and $\rho^L_{\alpha \beta}(\epsilon)$
 where, respectively, defined at (\ref{rho}) and (\ref{eq:rhoalphabeta}).

An alternative route to derive the coefficients (\ref{coef}) is by expanding the scattering matrix \cite{bode-2011,aleiner-2002,mobu,lilimos}
${\mathcal S}$ to leading order in $\dot{X}$,
\begin{equation}
\label{Sadexp}
{\cal S}(\epsilon,t) \simeq S(\epsilon,X(t))+  \dot{ X}(t) A(\epsilon,X(t))\,,
\end{equation}
where the strictly adiabatic S-matrix is given by
\begin{equation}
\label{FrozenS}
  S(\epsilon,X) = 1 - 2\pi \I W(\epsilon) G^{f,R}(\epsilon,X) W^\dagger(\epsilon)\,.
\end{equation}
We have defined
\begin{equation} 
W= \sqrt{\Gamma_L/\pi} \, \frac{\sigma^{0}+\sigma^{3}}{2}+ \sqrt{\Gamma_R/\pi}\, \frac{\sigma^{0}-\sigma^{3}}{2}\,,
\end{equation}
where $\sigma_{\mu}, \mu= 0, \ldots, 3$ denote the Pauli matrices in the two-site basis defined by the two quantum dots.
The A-matrices also take a
simple form for this model and is given by 
\begin{equation}\label{A2H2}
A(\epsilon,X)=-\I\lambda\,\frac{\sqrt{\Gamma_L\Gamma_R}}{g^2}\,\sigma^2\,.
\end{equation}

%%%%%%%%%%%%%%%%%%%%%%%%%%%%%%%%%%%%%%%%%%%%%%%%%%%%


\begin{thebibliography}{15}
\expandafter\ifx\csname natexlab\endcsname\relax\def\natexlab#1{#1}\fi
\expandafter\ifx\csname bibnamefont\endcsname\relax
  \def\bibnamefont#1{#1}\fi
\expandafter\ifx\csname bibfnamefont\endcsname\relax
  \def\bibfnamefont#1{#1}\fi
\expandafter\ifx\csname citenamefont\endcsname\relax
  \def\citenamefont#1{#1}\fi
\expandafter\ifx\csname url\endcsname\relax
  \def\url#1{\texttt{#1}}\fi
\expandafter\ifx\csname urlprefix\endcsname\relax\def\urlprefix{URL }\fi
\providecommand{\bibinfo}[2]{#2}
\providecommand{\eprint}[2][]{\url{#2}}

\bibitem{Safavi2012}A. H. Safavi-Naeini, J. Chan, J. T. Hill, T. P. Mayer Alegre, A. Krause, and O. Painter
Phys. Rev. Lett. {\bf 108}, 033602 (2012); A. Clerk, Physics {\bf 5}, 8 (2012).

\bibitem{mar-gir}F. Marquardt and S. Girvin, Physics {\bf 2}, 40 (2009); references therein. 

  
\bibitem[{\citenamefont{Zippilli et~al.}(2009)\citenamefont{Zippilli, Morigi,
  and Bachtold}}]{zippilli-2009}
\bibinfo{author}{\bibfnamefont{S.}~\bibnamefont{Zippilli}},
  \bibinfo{author}{\bibfnamefont{G.}~\bibnamefont{Morigi}}, \bibnamefont{and}
  \bibinfo{author}{\bibfnamefont{A.}~\bibnamefont{Bachtold}},
  \bibinfo{journal}{Phys. Rev. Lett.} \textbf{\bibinfo{volume}{102}},
  \bibinfo{pages}{096804} (\bibinfo{year}{2009}).


\bibitem[{\citenamefont{Zippilli et~al.}(2010)\citenamefont{Zippilli, Bachtold,
  and Morigi}}]{zippilli-2010}
\bibinfo{author}{\bibfnamefont{S.}~\bibnamefont{Zippilli}},
  \bibinfo{author}{\bibfnamefont{A.}~\bibnamefont{Bachtold}}, \bibnamefont{and}
  \bibinfo{author}{\bibfnamefont{G.}~\bibnamefont{Morigi}},
  \bibinfo{journal}{Phys. Rev. B} \textbf{\bibinfo{volume}{81}},
  \bibinfo{pages}{205408} (\bibinfo{year}{2010}).




\bibitem{gorelik-2011} F. Santandrea, L. Y. Gorelik, R. I. Shekhter, and M. Jonson, Phys. Rev. Lett. {\bf 106}, 186803 (2011).

\bibitem[{\citenamefont{Pistolesi}(2009)}]{pistolesi-2009}
\bibinfo{author}{\bibfnamefont{F.}~\bibnamefont{Pistolesi}},
  \bibinfo{journal}{Jour. Low Temp. Phys.} \textbf{\bibinfo{volume}{154}},
  \bibinfo{pages}{199} (\bibinfo{year}{2009}).

\bibitem[{\citenamefont{Galperin et~al.}(2009)\citenamefont{Galperin, Saito,
  Balatsky, and Nitzan}}]{galperin-2009}
\bibinfo{author}{\bibfnamefont{M.}~\bibnamefont{Galperin}},
  \bibinfo{author}{\bibfnamefont{K.}~\bibnamefont{Saito}},
  \bibinfo{author}{\bibfnamefont{A.~V.} \bibnamefont{Balatsky}},
  \bibnamefont{and} \bibinfo{author}{\bibfnamefont{A.}~\bibnamefont{Nitzan}},
  \bibinfo{journal}{Phys. Rev. B} \textbf{\bibinfo{volume}{80}},
  \bibinfo{pages}{115427} (\bibinfo{year}{2009}); P.R.Schiff and A.Nitzan, Chemical Physics {\bf 375}, 399-402 (2010).
  
  \bibitem[{\citenamefont{McEniry et~al.}(2009)\citenamefont{McEniry, Todorov,
  and Dundas}}]{mceniry-2009}
\bibinfo{author}{\bibfnamefont{E.~J.} \bibnamefont{McEniry}},
  \bibinfo{author}{\bibfnamefont{T.~N.} \bibnamefont{Todorov}},
  \bibnamefont{and} \bibinfo{author}{\bibfnamefont{D.}~\bibnamefont{Dundas}},
  \bibinfo{journal}{J.Phys.: Condens. Matter} \textbf{\bibinfo{volume}{21}},
  \bibinfo{pages}{195304} (\bibinfo{year}{2009}).

  
\bibitem{galptef} M. Galperin, A. Nitzan, and M. A. Ratner, Phys. Rev. B {\bf 75}, 155312 (2007).



\bibitem[{\citenamefont{Naik et~al.}(2006)\citenamefont{Naik, Buu, LaHaye,
  Armour, Clerk, Blencowe, and Schwab}}]{naik-2006}
\bibinfo{author}{\bibfnamefont{A.}~\bibnamefont{Naik}},
  \bibinfo{author}{\bibfnamefont{O.}~\bibnamefont{Buu}},
  \bibinfo{author}{\bibfnamefont{M.~D.} \bibnamefont{LaHaye}},
  \bibinfo{author}{\bibfnamefont{A.~D.} \bibnamefont{Armour}},
  \bibinfo{author}{\bibfnamefont{A.~A.} \bibnamefont{Clerk}},
  \bibinfo{author}{\bibfnamefont{M.~P.} \bibnamefont{Blencowe}},
  \bibnamefont{and} \bibinfo{author}{\bibfnamefont{K.~C.}
  \bibnamefont{Schwab}}, \bibinfo{journal}{Nature}
  \textbf{\bibinfo{volume}{443}}, \bibinfo{pages}{193} (\bibinfo{year}{2006}).
  
  \bibitem[{\citenamefont{Prance et~al.}(2009)\citenamefont{Prance, Smith,
  Griffiths, Chorley, Anderson, Jones, Farrer, and Ritchie}}]{prance-2009}
\bibinfo{author}{\bibfnamefont{J.~R.} \bibnamefont{Prance}},
  \bibinfo{author}{\bibfnamefont{C.~G.} \bibnamefont{Smith}},
  \bibinfo{author}{\bibfnamefont{J.~P.} \bibnamefont{Griffiths}},
  \bibinfo{author}{\bibfnamefont{S.~J.} \bibnamefont{Chorley}},
  \bibinfo{author}{\bibfnamefont{D.}~\bibnamefont{Anderson}},
  \bibinfo{author}{\bibfnamefont{G.~A.~C.} \bibnamefont{Jones}},
  \bibinfo{author}{\bibfnamefont{I.}~\bibnamefont{Farrer}}, \bibnamefont{and}
  \bibinfo{author}{\bibfnamefont{D.~A.} \bibnamefont{Ritchie}},
  \bibinfo{journal}{Phys. Rev. Lett.} \textbf{\bibinfo{volume}{102}},
  \bibinfo{pages}{146602} (\bibinfo{year}{2009}).

\bibitem[{\citenamefont{{Muhonen} et~al.}(2012)\citenamefont{{Muhonen},
  {Meschke}, and {Pekola}}}]{muhonen-2012}
\bibinfo{author}{\bibfnamefont{J.~T.} \bibnamefont{{Muhonen}}},
  \bibinfo{author}{\bibfnamefont{M.}~\bibnamefont{{Meschke}}},
  \bibnamefont{and} \bibinfo{author}{\bibfnamefont{J.~P.}
  \bibnamefont{{Pekola}}}, \bibinfo{journal}{Reports on Progress in Physics}
  \textbf{\bibinfo{volume}{75}}, \bibinfo{pages}{046501}
  (\bibinfo{year}{2012}), \eprint{1203.5100}.

  \bibitem{phon}
C. Chamon, E. R. Mucciolo, L. Arrachea, and R. B. Capaz, Phys. Rev. Lett. {\bf 106}, 135504 (2011);
L. Arrachea, E. R. Mucciolo, C. Chamon, and R. B. Capaz, Phys. Rev. B {\bf 86}, 125424 (2012).

\bibitem{casati1} G. Benenti, K. Saito, and G. Casati, Phys. Rev. Lett. {\bf 106}, 230602 (2011)
  
\bibitem{casati2} G. Benenti, G. Casati, T. Prosen, and K. Saito, arXiv:1311.4430.
  
\bibitem{seifert} K. Brandner, K. Saito, and U. Seifert,  Phys. Rev. Lett. {\bf 110}, 070603 (2013)
  
\bibitem{janine} S. Juergens, F. Haupt, M. Moskalets, and J. Splettstoesser,
  Phys. Rev. B {\bf 87}, 245423 (2013);
  F. Haupt, M. Leijnse, H. L. Calvo, L. Classen, J. Splettstoesser, and M. R. Wegewijs,  arXiv:1306.4343.
  
\bibitem{ora1} J.-H. Jiang, O. Entin-Wohlman, and Y. Imry, Phys. Rev. B {\bf 85} 075412 (2012).
  
\bibitem{ora2} O. Entin-Wohlman, J-H Jiang, Y. Imry,  arXiv:1309.5619.


\bibitem{chaste} J. Chaste,	 A. Eichler,	 J. Moser,	 G. Ceballos,	 R. Rurali, and A. Bachtold, Nature Nanotechnology {\bf 7}, 301 (2012).

\bibitem{fel-dot} A. Benyamini, A. Hamo, S. Viola Kusminskiy, F. von Oppen, S. Ilani,  Nature Phys. {\bf 10}, 151 (2014).
  
\bibitem{lu-2011} J-T L\"u, P. Hedegard, and M. Brandbyge, Phys. Rev. Lett. {\bf 107}, 046801 (2011).
  
 \bibitem[{\citenamefont{Koch et~al.}(2006)\citenamefont{Koch, von Oppen, and
  Andreev}}]{koch-2006}
\bibinfo{author}{\bibfnamefont{J.}~\bibnamefont{Koch}},
  \bibinfo{author}{\bibfnamefont{F.}~\bibnamefont{von Oppen}},
  \bibnamefont{and} \bibinfo{author}{\bibfnamefont{A.~V.}
  \bibnamefont{Andreev}}, \bibinfo{journal}{Phys. Rev. B}
  \textbf{\bibinfo{volume}{74}}, \bibinfo{pages}{205438}
  (\bibinfo{year}{2006}).

\bibitem{pisto-class}F. Pistolesi, Ya. M. Blanter, and I. Martin, Phys. Rev. B {\bf 78}, 085127 (2008).

\bibitem[{\citenamefont{Bode et~al.}(2011)\citenamefont{Bode, {Viola
  Kusminskiy}, Egger, and von Oppen}}]{bode-2011}
\bibinfo{author}{\bibfnamefont{N.}~\bibnamefont{Bode}},
  \bibinfo{author}{\bibfnamefont{S.}~\bibnamefont{{Viola Kusminskiy}}},
  \bibinfo{author}{\bibfnamefont{R.}~\bibnamefont{Egger}}, \bibnamefont{and}
  \bibinfo{author}{\bibfnamefont{F.}~\bibnamefont{von Oppen}},
  \bibinfo{journal}{Phys. Rev. Lett.} \textbf{\bibinfo{volume}{107}},
  \bibinfo{pages}{036804} (\bibinfo{year}{2011}).

\bibitem[{\citenamefont{Bode et~al.}(2012)\citenamefont{Bode, {Viola
  Kusminskiy}, Egger, and von Oppen}}]{bode-2012}
\bibinfo{author}{\bibfnamefont{N.}~\bibnamefont{Bode}},
  \bibinfo{author}{\bibfnamefont{S.}~\bibnamefont{{Viola Kusminskiy}}},
  \bibinfo{author}{\bibfnamefont{R.}~\bibnamefont{Egger}}, \bibnamefont{and}
  \bibinfo{author}{\bibfnamefont{F.}~\bibnamefont{von Oppen}},
  \bibinfo{journal}{Beilstein J. Nanotechnol.} \textbf{\bibinfo{volume}{3}},
  \bibinfo{pages}{144} (\bibinfo{year}{2012}).

  
\bibitem{en-an} H. L. Engquist and P. W. Anderson, Phys. Rev. B {\bf 24}, 1151 (1981).

\bibitem{leto1} L.F. Cugliandolo, J. Kurchan, and L. Peliti, Phys. Rev. E 55, 3898 (1997).

\bibitem{leto2} F. Zamponi, F. Bonetto, L.F. Cugliandolo, anf J. Kurchan,  J.Stat.Mech. P09013 (2005).
  


\bibitem{hugo} H. Aita, L. Arrachea, C. Na\'on and E. Fradkin, Phys. Rev. B {\bf 88}, 085122  (2013).
  
\bibitem{lili} L. Arrachea and E. Fradkin,  Phys. Rev. B {\bf 84}, 235436 (2011).
  


\bibitem{eftem} For a review, see L. Cugliandolo, J. Phys. A {\bf 44}, 483001 (2011).

\bibitem{koch-2004} J. Koch, F. von Oppen, Y. Oreg, and E. Sela, Phys, Rev. B {\bf 70}, 195107 (2004).

\bibitem{note}Here, we assume $t_0\ll \Gamma$ for definiteness. The conclusions remain
valid when $t_0$ and $\Gamma$ become of the same order. 

\bibitem[{\citenamefont{Aleiner et~al.}(2002)\citenamefont{Aleiner, Brouwer,
  and Glazman}}]{aleiner-2002}
\bibinfo{author}{\bibfnamefont{I.~L.} \bibnamefont{Aleiner}},
  \bibinfo{author}{\bibfnamefont{P.~W.} \bibnamefont{Brouwer}},
  \bibnamefont{and} \bibinfo{author}{\bibfnamefont{L.~I.}
  \bibnamefont{Glazman}}, \bibinfo{journal}{Physics Reports}
  \textbf{\bibinfo{volume}{358}}, \bibinfo{pages}{309 } (\bibinfo{year}{2002}).

\bibitem{mobu} M. Moskalets and M. B\"uttiker, Phys. Rev. B {\bf 69}, 205316 (2004).

\bibitem{lilimos} L. Arrachea and M. Moskalets, Phys. Rev. B {\bf 74}, 245322 (2006).



\end{thebibliography}
\end{document}